\documentclass[epj]{svjour}
\usepackage{graphics}
\usepackage{hyperref}
\newcommand{\be}{\begin{eqnarray}}
	\newcommand{\ee}{\end{eqnarray}}

\usepackage{color}
\usepackage{cite}
\usepackage{epsfig}
\usepackage{graphics}
\usepackage{amssymb}
\usepackage{verbatim}  
\usepackage{booktabs}
\usepackage{amsmath,array}

\def\lsim{\raise0.3ex\hbox{$<$\kern-0.75em\raise-1.1ex\hbox{$\sim$}}}
\def\gsim{\raise0.3ex\hbox{$>$\kern-0.75em\raise-1.1ex\hbox{$\sim$}}}
\begin{document}

\title{Fluctuations and correlations of net baryon number, electric charge and strangeness in a background magnetic field}

\author{H.-T. Ding\inst{1}\thanks{hengtong.ding@mail.ccnu.edu.cn}, S.-T. Li\inst{2,1}\thanks{stli@impcas.ac.cn}, Q. Shi\inst{1}\thanks{qi-shi@mails.ccnu.edu.cn}\and X.-D. Wang\inst{1}\thanks{xiaodanwang@mails.ccnu.edu.cn}
}                     
%
%
\institute{Key Laboratory of Quark \& Lepton Physics (MOE) and Institute of
	Particle Physics, \\Central China Normal University, Wuhan 430079, China \and Institute of Modern Physics, Chinese Academy of Sciences, Lanzhou 730000, China}
%
%
\abstract{
We present results on the second-order fluctuations of and correlations among net baryon number, electric
charge and strangeness in (2+1)-flavor lattice QCD in the presence of a background magnetic field.
Simulations are performed using the
tree-level improved gauge action and the highly improved staggered quark
(HISQ) action with a fixed scale approach ($a\simeq$ 0.117 fm). The light quark mass is set to be 1/10 of the physical strange quark mass and the corresponding pion mass is about 220 MeV at vanishing magnetic field. Simulations are performed on $32^3\times N_\tau$ lattices with 9 values of $N_\tau$ varying from 96 to 6 corresponding to temperatures ranging from zero up to 281 MeV. The magnetic field strength $eB$ is simulated with 15 different values up to $\sim$2.5 GeV$^2$ at each nonzero temperature. We find that quadratic fluctuations and correlations do not show any singular behavior at zero temperature in the current window of $eB$ while they develop peaked structures at nonzero temperatures as $eB$ grows. By comparing the electric charge-related fluctuations and correlations with hadron resonance gas model calculations and ideal gas limits we find that the changes in degrees of freedom start at lower temperatures in stronger magnetic fields. Significant effects induced by magnetic fields on the isospin symmetry and ratios of net baryon number and baryon-strangeness correlation to strangeness fluctuation are observed, which could be useful for probing the existence of a magnetic field in heavy-ion collision experiments. 
} 

\authorrunning{H.-T. Ding, S.-T. Li, Q. Shi, X.-D. Wang}
\titlerunning{Fluctuations and correlations of B, Q \& S in a background magnetic field}
\maketitle

\section{Introduction}
QCD phase structure in the nonzero magnetic fields has attracted intensive interest recently as the strong magnetic field is expected to be produced in the early stage of peripheral heavy-ion collisions~\cite{Kharzeev:2007jp,Skokov:2009qp,Deng:2012pc}, early universe~\cite{Vachaspati:1991nm} and magnetars~\cite{enqvist1993primordial}.  In heavy-ion collisions given that the magnetic field lives sufficiently long the chiral magnetic effect shall be manifested in the experimental observations ~\cite{Kharzeev:2015znc,Kharzeev:2020jxw}. The lifetime of the magnetic field strongly depends on the electrical conductivity of the medium, whose determination, however, is difficult due to the inverse problem in the first principle computations~\cite{Astrakhantsev:2019zkr,Ding:2016hua,Ding:2010ga,Aarts:2007wj}. Many efforts have been made to search for the signal of a magnetic field in the heavy-ion collision experiments. Recent observations of differences of direct flows between $D^0$ and $\bar{D}^0$~\cite{Adam:2019wnk,Acharya:2019ijj} and the broadening of transverse momentum distribution of dileptons produced through photon fusion processes~\cite{Adam:2018tdm,Aaboud:2018eph} in heavy-ion collisions might indicate the possible existence of a magnetic field in the deconfined quark-gluon plasma phase. On the other hand, in the presence of an external magnetic field, up and down quarks cannot be considered as isospin symmetric anymore due to their different electric charges. The magnitude of isospin symmetry breaking manifested in the difference between up and down quark chiral condensates has been computed from lattice QCD~\cite{Bali:2012zg,Ding:2020hxw}, however, the chiral condensates are surely not measurable in experiments.

Based on lattice QCD studies it is well-known that a strong magnetic field can bring interesting effects on QCD thermodynamics~\cite{Bali:2014kia}, phase diagram~\cite{Bali:2011qj,Ding:2020inp}, transport properties~\cite{Astrakhantsev:2019zkr} as well as hadron spectroscopy~\cite{Bonati:2015dka,Bali:2017ian,Ding:2020hxw,Endrodi:2019whh}. In particular the inverse magnetic catalysis with a reduction of chiral crossover transition temperature $T_{pc}$ in external magnetic fields~\cite{Ilgenfritz:2013ara,Bornyakov:2013eya,Bali:2014kia,Tomiya:2019nym} have triggered a lot of interests~\cite{Cao:2021rwx,Shovkovy:2012zn,Andersen:2014xxa,DElia:2011koc,Kojo:2012js,Bruckmann:2013oba,Fukushima:2012kc,Ferreira:2014kpa,Yu:2014sla,Feng:2015qpi,Li:2019nzj,Mao:2016lsr,Gursoy:2016ofp,Xu:2020yag}. However, much less is known about the details on changes in degrees of freedom in QCD with an external magnetic field.

Fluctuations of and correlations among net baryon number (B),  strangeness (S), and electric charge (Q) have been very useful to probe the changes of degrees of freedom at zero magnetic fields and the QCD phase structure, as they are both theoretically computable and experimentally measurable~\cite{Luo:2017faz,Ding:2015ona}. They have been extensively employed to study the changes in the degree of freedom in the system~\cite{Asakawa:2000wh,Jeon:2000wg,Koch:2005vg,Ding:2015fca,Bazavov:2013dta,Bazavov:2014yba} and probe the critical end point~\cite{Stephanov:1998dy,Stephanov:1999zu,Friman:2011pf,Luo:2017faz,Adam:2020unf,Fu:2021oaw,Ding:2020rtq,Rustamov:2020ekv}. For instance, the ratio of 4th order fluctuation of baryon number to the 2nd one~\cite{Bazavov:2017dus}, and the ratio of baryon-strangeness correlation to quadratic strangeness fluctuations~\cite{Koch:2005vg} are useful to probe the deconfinement properties of the QCD transition~\cite{Bazavov:2017dus}. 
On the other hand, fluctuations and correlations of B, Q and S in nonzero magnetic fields are much less explored. Most of the studies in the literature are based on the hadron resonance gas model~\cite{Fukushima:2016vix,Ferreira:2018pux,Bhattacharyya:2015pra,Kadam:2019rzo} and Polyakov-Nambu-Jona-Lasinio model~\cite{Fu:2013ica}, and there do not exist any studies based on lattice QCD.  It has been found from e.g. Ref.~\cite{Fukushima:2016vix} the quadratic electric charge fluctuation is largely enhanced in particular at high baryon density based on studies using the hadron resonance gas model. The hadron resonance gas model based on Dashen-Ma-Bernstein theorem~\cite{Dashen:1969ep} is supposed to describe QCD only at low temperature where QCD is well approximated by the non-interacting hadron resonance mass. Thus it would be useful to have first principle computations on these quantities.

In this paper we will present a first lattice QCD computation on the quadratic fluctuations and correlations of net baryon number, electric charge and strangeness in the presence of constant external magnetic fields. We will show that the isospin symmetry breaking can be directly observed in certain combinations of fluctuations and correlations of B, Q and S. We will also compare our results with those obtained from the hadron resonance gas model at low temperatures and the high-temperature ideal gas limit. Connections for probing the magnetic field in the late stage of heavy-ion collision are also discussed. The computation is based on lattice QCD simulations using highly improved staggered fermions at a single lattice spacing $a\simeq$ 0.117 fm with pion mass about 220 MeV at vanishing magnetic field. We adopted a fixed scale approach to varying the temperature from zero up to $\sim$281 MeV, and the strength of magnetic fields $eB$ varies from 0 to $\sim 2.5$ GeV$^2$.

The paper is organized as follows. At the beginning of Section~\ref{sec:basics} we will give a basic definition of the quadratic fluctuations of and correlations among net baryon number, electric charge and strangeness, and in Section~\ref{sec:subHRG} we will give a brief description of the hadron resonance gas model, and show explicit formulae of the quadratic fluctuations and correlations of net baryon number, electric charge and strangeness in presence of an external magnetic field, and in Section~\ref{sec:subfree} we will then derive the quadratic fluctuations and correlations in the high-temperature free limit with nonzero $eB$. In Section~\ref{sec:lattice} we present details of our lattice setup. In Section~\ref{sec:resobs} we show temperature dependences of the fluctuations and correlations in strong magnetic fields, and in Section~\ref{sec:resiso} we present the magnitude of isospin symmetry breaking effects induced by the magnetic fields, and in Section~\ref{sec:resHRGfreePH} we compare our results to the hadron resonance gas model and the ideal gas limit and show ratios of fluctuations and correlations which could be investigated in heavy-ion experiments. Finally, we summarize our results in Section~\ref{sec:summary}.
Some preliminary results have been reported in proceedings~\cite{Ding:2020pao}.

\section{Fluctuations and correlations of conserved charges at nonzero magnetic field}
\label{sec:basics}
To calculate the fluctuations of conserved charges and their correlations in a thermal medium, the starting point is the pressure $p$ expressed in terms of the logarithm of partition function $Z$ as follows
\begin{equation}
	\frac{p}{T^4} \equiv \frac{1}{VT^3}\ln Z(V,T,\mu_{\rm B},\mu_{\rm S},\mu_{\rm Q}) \; ,
	\label{pressure}
\end{equation}
where the baryon ($\mu_{\rm B}$), strangeness ($\mu_{\rm S}$) and electric charge ($\mu_{\rm Q}$) chemical potentials have following relations with the quark chemical potentials $\mu_{ u }$, $\mu_{ d }$ and $\mu_{ s }$,
\begin{align}
\begin{split}
	\mu_u&=\frac{1}{3}\mu_{\rm B} + \frac{2}{3}\mu_{\rm Q} \; , \\
	\mu_d&=\frac{1}{3}\mu_{\rm B} - \frac{1}{3}\mu_{\rm Q} \; , \\
	\mu_s&=\frac{1}{3}\mu_{\rm B} - \frac{1}{3}\mu_{\rm Q} - \mu_{\rm S} \; .
	\end{split}
	\label{potential}
\end{align}

The fluctuations of the conserved charges and their correlations can be obtained by taking the derivatives of pressure with respect to the chemical potentials from lattice calculation evaluated at zero chemical potentials~\cite{Bazavov:2012jq},
\begin{equation}
	\begin{split}
		&\hat{\chi} _ { i j k } ^ { u d s }  = \frac { \partial ^ { i + j + k } p / T ^ { 4 } } { \partial \left( \mu _ { u } / T \right) ^ { i } \partial \left( \mu _ { d } / T \right) ^ { j } \partial \left( \mu _ { s } / T \right) ^ { k } }\Bigg{|}_{\mu_{u,d,s}=0},\\
		&\hat{\chi} _ { i j k } ^ { \rm B Q S }  = \frac { \partial ^ { i + j + k } p / T ^ { 4 } } { \partial \left( \mu _ {\rm  B } / T \right) ^ { i } \partial \left( \mu _ { \rm Q } / T \right) ^ { j } \partial \left( \mu _ {\rm  S } / T \right) ^ { k } }\Bigg{|}_{\mu_{\rm B, Q, S}=0}.
	\end{split}
\end{equation}
Here in our study we focus on the computation of quadratic fluctuations and correlations, i.e. $i+j+k=2$. The expressions of quadratic fluctuations $\hat{\chi} _ { i j k } ^ { \rm B Q S }$ in terms of $\hat{\chi} _ { i j k } ^ { u d s }$ can be easily obtained via Eq.~\ref{potential}, and the explicit forms can be found in e.g. Ref.~\cite{Petreczky:2012rq}. Here for the discussion of isospin symmetry breaking we list the expression of $\chi_2^{u,d}$ in terms of fluctuations and correlations of ${\rm B,Q,S}$ as follows
\begin{align}
\begin{split}
    \chi_2^u &= \chi_2^{\rm B}+\chi_2^{\rm Q} + 2\chi_{11}^{\rm BQ} ,\\
    \chi_2^{d} &= 4\chi_2^{\rm B}+\chi_2^{\rm Q} +\chi_2^{\rm S} - 4\chi_{11}^{\rm BQ} -2\chi_{11}^{\rm QS}+4\chi_{11}^{BS}.
    \end{split}
    \label{eq:chiudBQS}
\end{align}

\subsection{Hadron resonance gas model}
\label{sec:subHRG}

In the hadron resonance gas (HRG) model, the pressure arising from charged and neutral particles in the presence of a magnetic field can be expressed as follows (see Refs. \cite{Bazavov:2012jq} for the case of $eB=0$ and \cite{Fukushima:2016vix,Endrodi:2013cs,Bhattacharyya:2015pra} for the relation at $eB \neq 0$),
\begin{small}
	\begin{align}
	&p_c^{\mathrm{M/B}} = \mp \frac{|q_i| B T}{2\pi^2} \sum_{s_z=-s_i}^{s_i} \sum_{l=0}^\infty \int_0^\infty {\rm d} p_z \ln\left[ 1 \mp e^{-(E_c - \mu_i)/T} \right], \\
	&p_n^{\mathrm{M/B}} = \mp \frac{d_i T}{2\pi^2} \int_0^\infty {\rm d}p |\vec{p}|^2 \ln\left[ 1 \mp e^{-(E_n - \mu_i)/T} \right],
	\end{align}
\end{small}
 respectively.
Here
$E_c = \sqrt{p_z^2 + m_i^2 + 2|q_i|B (l +1/2 - s_z)}$ and $E_n = \sqrt{m_i^2 + |\vec{p}|^2}$
denote the energy levels of the charged and neutral particles with momentum $\vec{p} = \linebreak (p_x, p_y, p_z)$, respectively. $q_i$, $m_i$, $s_i$ and $d_i$ are the charge, mass, spin and degeneracy factor of the particle $i$, $B$ is the magnitude of magnetic field pointing along the $z$ direction, $l$ denotes the Landau levels, and $\mu_i=\mu_{\rm B}{\rm B}_i+\mu_{\rm Q}{\rm Q}_i+\mu_{\rm S}{\rm S}_i$ with ${\rm B}_i$, ${\rm Q}_i$ and ${\rm S}_i$ the baryon number, charge and strangeness of the particle $i$, respectively. Here ``+" in ``$\mp$"corresponds to the case for mesons ($s_i$ is integer) while ``$-$" for baryons ($s_i$ is half-integer).

After integrating out the momentum  we arrive at the analytical expressions of the pressure~\footnote{Here we neglect the term arising from the vacuum energy, as which receives no contributions to the fluctuations and correlations of B, Q and S.},
\begin{small}
	\begin{align}
	&\frac{p_c^{M/B}}{T^4} = \frac{|q_i| B}{2\pi^2T^3} \sum_{s_z=-s_i}^{s_i} \sum_{l=0}^\infty \varepsilon_0 \sum_{k=1}^\infty (\pm 1)^{k+1} \frac{e^{k\mu_i/T}}{k} {\rm K}_1 \left(\frac{k\varepsilon_0}{T}\right), \label{eq: HRGp} \\
	&\frac{p_n^{M/B}}{T^4} = \frac{d_i m_i^2}{2(\pi T)^2} \sum_{k=1}^\infty (\pm 1)^{k+1} \frac{e^{k\mu_i/T}}{k^2} {\rm K}_2 \left(\frac{km_i}{T}\right)\,,
	\end{align}
\end{small}
where 
\begin{align}
\varepsilon_0 = \sqrt{m_i^2+2|q_i|B(l+1/2-s_z)}
\label{eq:Elevel}
\end{align} 
are the energy levels of charged particles with $p_z = 0$, and $k$ is the sum index in the Taylor expansion series. $K_1$ and $K_2$ are the first-order and second-order modified Bessel functions of the second kind, respectively. For the charged particle in the presence of a magnetic field, by taking derivatives of Eq.~\ref{eq: HRGp} with respect to chemical potentials of conserved charges and then setting $\vec{\mu} = (\mu_{\rm B},\mu_{\rm Q},\mu_{\rm S})=0$, one arrives at
	\begin{align}
	\begin{split}
	&\chi_2^X = \frac{B}{2\pi^2 T} \sum_i |q_i| X_i^2 \sum_{s_z=-s_i}^{s_i} \sum_{l=0}^\infty f (\varepsilon_0) , \\
	&\chi_{11}^{XY} = \frac{B}{2\pi^2 T} \sum_i |q_i| X_iY_i \sum_{s_z=-s_i}^{s_i} \sum_{l=0}^\infty f (\varepsilon_0) \,,
	\end{split}
	\label{eq:HRGresults}
	\end{align}
with $X, Y = {\rm B, Q, S}$ and $f(\varepsilon_0)=\varepsilon_0 \sum_{k=1}^\infty (\pm 1)^{k+1} k {\rm K}_1\left(\frac{k\varepsilon_0}{T}\right)$. We remark here that Eq.~\ref{eq:Elevel} only holds true in the case that charged hadrons can still be considered as point-like particles in the magnetic field. As presented in Ref.~\cite{Ding:2020hxw} energy levels of both $\pi^+$ and $K^-$ deviate from Eq.~\ref{eq:Elevel} at $eB\gtrsim 0.3$ GeV$^2$. For simplicity we will thus consider only the case when $eB\lesssim 0.3$ GeV$^2$ in our current study. In our HRG model treatment, we have incorporated all the hadrons listed in the particle data group (PDG) \cite{Tanabashi:2018oca} up to the mass of 2.5 GeV.

\subsection{Ideal gas limit}
\label{sec:subfree}
In the high-temperature (free) limit, by following textbooks~\cite{Kapusta:2006pm,Laine:2016hma}, pressure of QCD with three massless flavor quarks in the nonzero magnetic field can be derived and expressed as follows
\begin{align}
\frac{p}{T^4} =\frac{8\pi^2}{45} + \sum_{f=u,d,s} \frac{3|q_f|B}{\pi^2T^2} \left[  \frac{\pi^2}{12} + \frac{\hat{\mu}_f^2}{4}+ p_f(B) \right],
\label{eq:freep_uds}
\end{align}
where 
\begin{align}
p_f(B)=2\frac{\sqrt{2|q_f|B}}{T}\sum_{l=1}^{\infty} \sqrt{l}\sum_{k=1}^\infty \frac{(-1)^{k+1}}{k} &\cosh\left(k\hat{\mu}_f\right) \times \nonumber \\&{\rm K}_1\left(\frac{k\sqrt{2|q_f|Bl}}{T}\right),
\end{align}
$q_f$ denotes the electric charge of a quark flavor $f$ and $\hat{\mu}_f\equiv\mu_f/T$.  We remark here that unlike the case at $eB=0$ the pressure of a free massless three flavor quark gas with $eB\ne0$ receives contributions from terms beyond $\mathcal{O}(\mu_f^4)$, and fluctuations and correlations of quarks which are higher than the 4th order thus could survive in the magnetized free gas.
Here we focus on the 2nd order fluctuations and correlations. By taking derivatives of Eq.~\ref{eq:freep_uds} with respect to quark chemical potentials and then setting $\mu_{u, d, s} = 0$, one can get
\begin{align}
&\frac{\chi_2^u}{eB} = \frac{4}{\pi^2}\left(\frac{1}{4} +  \hat{b} \sum_{l=1}^{\infty} \sqrt{2\l} \sum_{k=1}^\infty (-1)^{k+1}k \,{\rm K}_1\left(k\,\hat{b}\,\sqrt{2l}\right)\right)\,, \\
&\frac{\chi_2^{d,s}}{eB} =\frac{2}{\pi^2}\left( \frac{1}{4} + \hat{b} \sum_{l=1}^{\infty} \sqrt{l} \sum_{k=1}^\infty (-1)^{k+1}k \,{\rm K}_1\left(k\,\hat{b}\,\sqrt{l}\right)\right)\,,	\\
&\chi_{11}^{ud} = \chi_{11}^{us} = \chi_{11}^{ds} = 0.
\end{align}
Here we use $\hat{b}\equiv\sqrt{2eB/3}/T$ for brevity. Using Eq.~\ref{potential}, the second-order fluctuations of and correlations among net baryon number, electric charge and strangeness in the high-temperature limit can then be expressed as follows
\begin{align}
\label{eq:freeB}
\frac{\chi_2^{\rm B}}{eB} = \frac{4}{9\pi^2} \Bigg(\frac{1}{2} + &\hat{b} \sum_{l=1}^{\infty} \sqrt{l} \sum_{k=1}^\infty (-1)^{k+1}k\nonumber\\
\times &\left[\sqrt{2}\,{\rm K}_1\left(k\,\hat{b}\,\sqrt{2l}\right) + {\rm K}_1\left(k\,\hat{b}\,\sqrt{l}\right) \right]\Bigg)\,\\
\frac{\chi_2^{\rm Q}}{eB} = \frac{4}{9\pi^2} \Bigg(\frac{5}{4} + &\hat{b} \sum_{l=1}^{\infty} \sqrt{l} \sum_{k=1}^\infty (-1)^{k+1}k\nonumber \label{eq:free2Q}\\
\times &\left[4\sqrt{2}\,{\rm K}_1\left(k\,\hat{b}\,\sqrt{2l}\right) + {\rm K}_1\left(k\,\hat{b}\,\sqrt{l}\right) \right]\Bigg),\
\end{align}
\begin{align}
 \chi_2^{\rm S} = \chi_2^s\,,\quad&\\
\frac{\chi_{11}^{\rm BQ}}{eB} = \frac{4}{9\pi^2} \Bigg(\frac{1}{4} + &\hat{b} \sum_{l=1}^{\infty} \sqrt{l} \sum_{k=1}^\infty (-1)^{k+1}k\nonumber \label{eq:free11BQ} \\
\times &\left[2\sqrt{2}\,{\rm K}_1\left(k\,\hat{b}\,\sqrt{2l}\right) - {\rm K}_1\left(k\,\hat{b}\,\sqrt{l}\right) \right]\Bigg),\\
\frac{\chi_{11}^{\rm QS}}{eB} = \frac{2}{3\pi^2}  \Bigg(\frac{1}{4} +& \hat{b} \sum_{l=1}^{\infty} \sqrt{l} \sum_{k=1}^\infty (-1)^{k+1}k {\rm K}_1\left(k\hat{b}\sqrt{l}\right)\Bigg).\label{eq:free11QS}\\
\chi_{11}^{\rm BS}=-\chi_{11}^{QS}  \,.\quad&
\label{eq:freeBQS}
\end{align}
 It can be observed that all these fluctuations and correlations divided by $eB$ scale with $\sqrt{eB}/T$.
 From above relations it can also be found that 
\begin{align}
\chi_{11}^{\rm BS}/\chi_2^{\rm S} = -\chi_{11}^{\rm QS}/\chi_2^{\rm S}= -\frac{1}{3}\,,
\label{eq:free2}
\end{align}
which is the same as the case at zero magnetic field. The following relations also hold true at both $eB=0$ and $eB\neq0$ in the free limit
\begin{align}
    \chi_2^d=\chi_2^s,~~~\chi_{11}^{ud}=\chi_{11}^{us}=\chi_{11}^{us}=0.
\end{align}

In Table~\ref{tab:largeBfree} we also list the values of the above quantities in the case of $\sqrt{eB}/T$ going to infinity in the free limit. 
\begin{table}[!htpb]
	\centering
	\begin{tabular}{c|c}
		\toprule[1.5pt]
		Quantity & Value \\ 
		\hline
		$\chi_2^u/eB$ & $1/\pi^2$ \\
		$\chi_2^{d/s/S}/eB$ & $1/(2\pi^2)$ \\
		$\chi_{11}^{ud}/eB$ = $\chi_{11}^{us}/eB$ = 		$\chi_{11}^{ds}/eB$=0& 0 \\		
		$\chi_2^{\rm B}/eB$ & $2/(9\pi^2) $ \\
		$\chi_2^{\rm Q}/eB$ & $5/(9\pi^2) $ \\
		$\chi_{11}^{\rm BQ}/eB$ & $1/(9\pi^2)$ \\
		$\chi_{11}^{\rm QS}/eB = -\chi_{11}^{\rm BS}/eB = \chi_2^{\rm S}/3eB$ & $1/(6\pi^2)$\\
		\bottomrule[1.5pt]
	\end{tabular} 
	\caption{The second order fluctuations and correlations of B, Q and S ($u$, $d$ and $s$) divided by the magnetic field strength $eB$ in the ideal gas limit with $\sqrt{eB}/T$ going to infinity.}
	\label{tab:largeBfree}
\end{table}
For comparison we also list here the high-temperature limits of various fluctuations and correlations of B, Q and S for massless three flavor quark gas at $eB=0$~\cite{Bazavov:2012jq}
\begin{align}
\begin{split}
&\chi_{2}^{\rm B} = \chi_{11}^{\rm QS} = -\chi_{11}^{\rm BS} =\chi_2^{\rm Q}/2=\chi_2^{\rm S}/3= 1/3,\\
&\chi_{11}^{\rm BQ}=0\,.
\end{split}
\end{align}

\section{Lattice setup}
\label{sec:lattice}
The highly improved staggered quarks (HISQ)~\cite{Follana:2006rc} and a tree-level improved Symanzik gauge action, which have been extensively used by the HotQCD collaboration~\cite{Bazavov:2019www}, were adopted in our current lattice simulations of $N_f=2+1$ QCD in nonzero magnetic fields. The magnetic field is introduced along the $z$ direction, and is described by a fixed factor $u_{\mu}(n)$ of the U(1) field. $u_{\mu}(n)$ can be expressed as follows in the Landau gauge
 \cite{AlHashimi:2008hr,Bali:2011qj},
\begin{eqnarray} \label{eq:BC_extB}
u_x(n_x,n_y,n_z,n_\tau)&=&
\begin{cases}
\exp[-iq a^2 B N_x n_y] \;\;&(n_x= N_x-1)\\
1 \;\;&(\text{otherwise})\\
\end{cases}\notag\\
u_y(n_x,n_y,n_z,n_\tau)&=&\exp[iq a^2 B  n_x],\label{eq:def_mag_u} \notag\\
u_z(n_x,n_y,n_z,n_\tau)&=&u_t(n_x,n_y,n_z,n_\tau)=1.
\end{eqnarray}
Here the lattice size is denoted as $(N_x,\; N_y,\; N_z,\; N_\tau)$ and coordinates as $n_\mu=0,\cdots, N_\mu-1$ ($\mu = x,\; y,\; z,\; \tau$).
To satisfy the quantization for all the quarks in the system, the greatest common divisor of the electric charge of all the quarks, i.e. $|q_d|=|q_s|=e/3$ with $e$ the elementary electric charge, is chosen in our simulation. In practice, the strength of the magnetic field $eB$ is expressed as follows
\begin{equation}
eB= \frac{6 \pi N_{b}}{N_{x} N_{y}} a^{-2},
\label{eq:eBdef}
\end{equation}
where $N_b \in \mathbf{Z}$ is the number of magnetic fluxes through a unit area in the $x$-$y$ plane. The periodic boundary
condition for U(1) links is applied for all directions except for
the $x$-direction, as shown in Eq.\ref{eq:BC_extB}. As limited by the boundary condition, $N_b$ is constrained in the range of $0\leq N_b<\frac{N_xN_y}{4}$. In our study $N_\sigma\equiv N_x=N_y=N_z$. Details about the implementation of magnetic fields in the lattice QCD simulations using the HISQ action can be found in Ref.~\cite{Ding:2020hxw}, where similar procedures were adopted at zero temperature.

\begin{table}[htp!]
	\begin{minipage}{0.6\hsize}
		\centering
		\begin{tabular}{|cc|cc||ccc|}
			
			\hline
			$N_{b}$ & $eB$ [GeV$^2$] &$N_{b}$ & $eB$ [GeV$^2$] & $N_{\tau}$ & $T$ [MeV] & \# conf.\\
			\hline
			0 & 0 & 16 & 0.836 &6 & 280.9 	 & $\mathcal{O}(4000)$\\
			1 & 0.052 & 	20 & 1.045 &8 & 210.8 & $\mathcal{O}(5000)$\\			
			2 & 0.104 & 	24 & 1.255 &10 & 168.5 & $\mathcal{O}(5000)$\\			   			
			3 & 0.157 &	32 & 1.673 & 12 & 140.4 & $\mathcal{O}(5000)$\\	
			4 & 0.209 & 40 & 2.09 &14 & 120.4 & $\mathcal{O}(5000)$\\			   			   			   					   			   			
			6 & 0.314 & 	48 & 2.510 &16 &  105.3 & $\mathcal{O}(6000)$\\			   	
			8 & 0.418 &      -& - &18  &  93.6& $\mathcal{O}(6000)$\\ 
			10 & 0.523 & 	-& - &24  &  70.2& $\mathcal{O}(1000)$\\    			
			12 & 0.627 &	- & - & 96  &  17.6& $\mathcal{O}(3000)$\\    			   			
			\hline
			
		\end{tabular}\\
		\vspace{0.1cm}
	\end{minipage}
	\caption{Statistics, values of $N_{b}$ and corresponding magnetic field strength $eB$, and values of $N_{\tau}$ and corresponding temperatures in the simulation. The lattice spacing is fixed to $a\simeq0.117$ fm ($a^{-1}\simeq 1.685$ GeV), pion mass at $eB=0$ is $M_\pi=220.61(6)$ MeV and the kaon decay constant is $f_K=112.50(2)$ MeV~\cite{Ding:2020hxw}.}
	\label{tab:setup}
\end{table}
In our lattice simulations, the strange quark mass is fixed to its physical value $m_{s}^{\rm phy}$ and the light quark masses are chosen to be $m_{s}^{\rm phy}/10$, which correspond to a Goldstone pion mass $m_{\pi }\simeq220$ MeV at zero magnetic field~\cite{Ding:2020hxw}. To perform simulations at nonzero temperature extending from our study at zero temperature~\cite{Ding:2020hxw}, we adopted a fixed scale approach, i.e. fixed lattice spacing $a\simeq0.117$ fm in our simulations. Variation of temperatures are obtained by varying the values of $N_\tau$ as $T=a^{-1}/N_\tau$. Values of $N_{\tau}$ are chosen from 96 to 6 corresponding to values of temperature ranging from zero temperature up to about 281 MeV as shown in Table.~\ref{tab:setup}.  The scale setting is adopted from the HotQCD collaboration~\cite{Bazavov:2019www}. For most of each fixed $N_{\tau}$, we have around 15 magnetic field flux $N_{b}$ values chosen from 0 to 48. These correspond to the magnetic field $eB$ ranging from 0 to $\sim$2.5 GeV$^2$ as shown in the Table.~\ref{tab:setup}~\cite{Ding:2020hxw}. To have small discretization errors for $B$, the magnetic field implemented in the lattice simulations should be small in lattice units, i.e. $aq_dB\ll 1$ or $N_b/N_\sigma^2\ll1$~\cite{Endrodi:2019zrl}. In our work, the largest number of magnetic fluxes $N_b^{max}=48$ resulting in $N_b^{max}/N_\sigma^2\approx5\%$. Thus the discretization errors for $B$ should be small. 

All configurations have been produced using the Rational Hybrid Monte Carlo (RHMC) algorithm and saved by every 5 time units. The number of saved configurations for each $N_b$ at each temperature is listed in Tab.~\ref{tab:setup}. The fluctuations and correlations of conserved charges at nonzero magnetic fields have been computed using the random noise vector method with 102 random vectors on each saved configuration.

We remark that the fixed scale approach is different from the commonly adopted approach used in e.g. Ref.~\cite{Bazavov:2020bjn,Ding:2019prx,Borsanyi:2020fev,Bali:2011qj,DElia:2018xwo} where lattice spacing $a$ varies at fixed $N_\tau$ to have different temperatures, and has also been adopted in $\linebreak$quenched QCD~\cite{Ding:2016hua} as well as full QCD~\cite{Taniguchi:2020mgg}. In the fixed scale approach we have the same value of $a^{-1}$ at various temperatures, and $eB$ thus only varies with $N_b$ (cf.~Eq.~\ref{eq:eBdef}). This is different from the commonly adopted approach, where interpolations of lattice data at different $T$ and $N_b$ are needed to have constant magnetic field strength in physical units (e.g. GeV$^2$) among different temperatures as $a$ varies with temperature~\cite{Bali:2011qj,DElia:2018xwo}. Comparing to the state-of-the-art lattice computation of fluctuations of conserved charges at zero magnetic field~\cite{Bazavov:2020bjn}, the lattice spacing adopted in our study is smaller than those on $N_\tau=6$ lattices with $T\lesssim 281$ MeV, $N_\tau=8$ lattices with $T\lesssim211$ MeV, $N_\tau=10$ lattices with $T\lesssim 169$ MeV, $N_\tau=12$ lattices with $T\lesssim140$ MeV and $N_\tau=16$ lattices with $T\lesssim105$ MeV.

\section{Results}
\label{sec:res}

\subsection{Fluctuations and correlations of net baryon number, electric charge and strangeness}

\begin{figure}[htp!]			
	\begin{center}
		\includegraphics[width=0.35\textwidth]{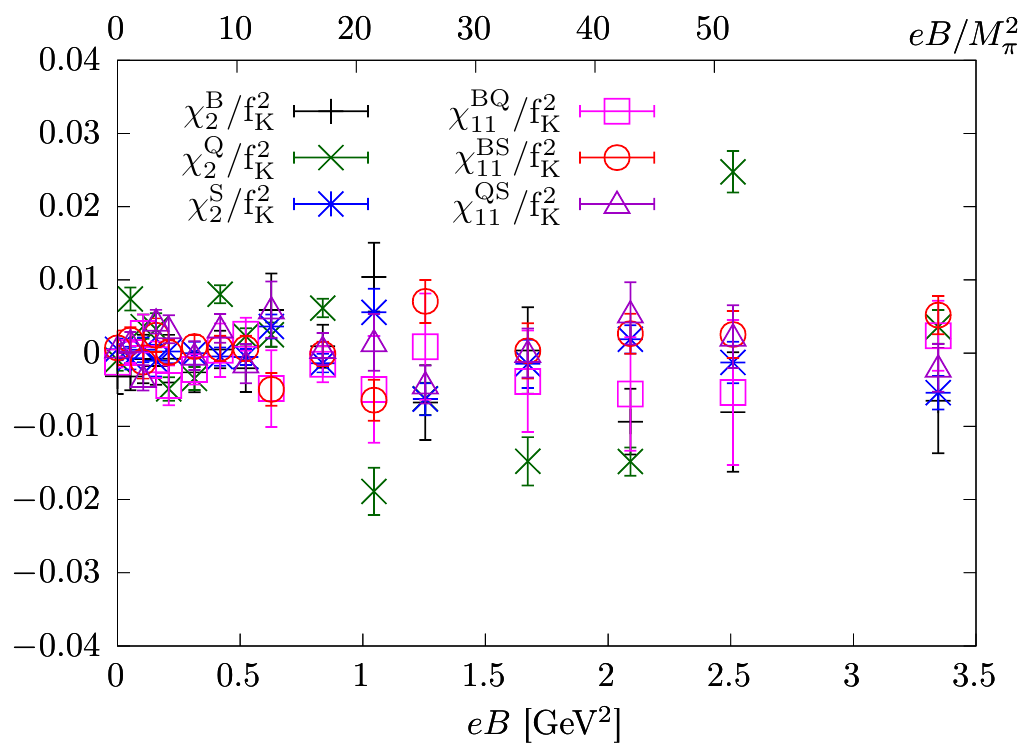}~					
	\end{center}
	\caption{$eB$ dependences of fluctuations of and correlations among conserved charges at $T$=0. Here kaon decay constant $f_K\simeq 112.5$ MeV obtained in the current lattice setup~\cite{Ding:2020hxw} is used to make quantities dimensionless. Hereafter $M_\pi$ located near the upper $x$-axis denotes the pion mass of 220 MeV at $eB=0$ in our lattice setup.}	
	\label{fig:fcT0}
\end{figure}

We start by showing the fluctuations of and correlations among conserved charges at zero temperature in Fig.~\ref{fig:fcT0}. It has been conjectured that there could be a superconducting phase induced by the strong magnetic field at zero temperature~\cite{Chernodub:2011mc}, which can be signaled by the condensation of vector meson $\rho$. As $\rho$ is a boson whose energy levels obey the Bose-Einstein distribution, if any vanishing energy level appears the fluctuations or correlations of quantum numbers receiving contributions from charged mesons would be divergent. However, as can be seen from Fig.~\ref{fig:fcT0} there is no divergent behavior in $\chi_2^{\rm Q}$ and all other fluctuations and correlations observed in the window of the magnetic field we studied. This provides a shred of indirect evidence that no superconducting phase exits at $eB\lesssim 3.5$ GeV$^2$, which is consistent with studies of hadron spectrum at zero temperature in quenched~\cite{Bali:2017ian} and full QCD~\cite{Ding:2020hxw}.
\label{sec:resobs}

\begin{figure*}[t]			
	\begin{center}
		\includegraphics[width=0.31\textwidth]{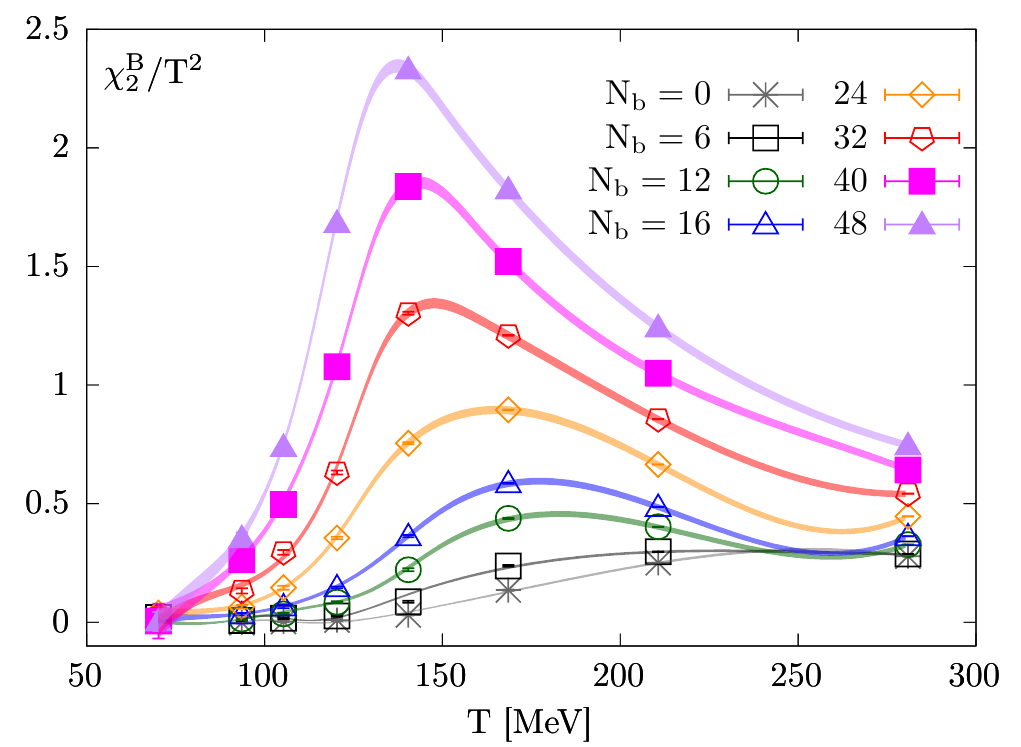}	
		\includegraphics[width=0.31\textwidth]{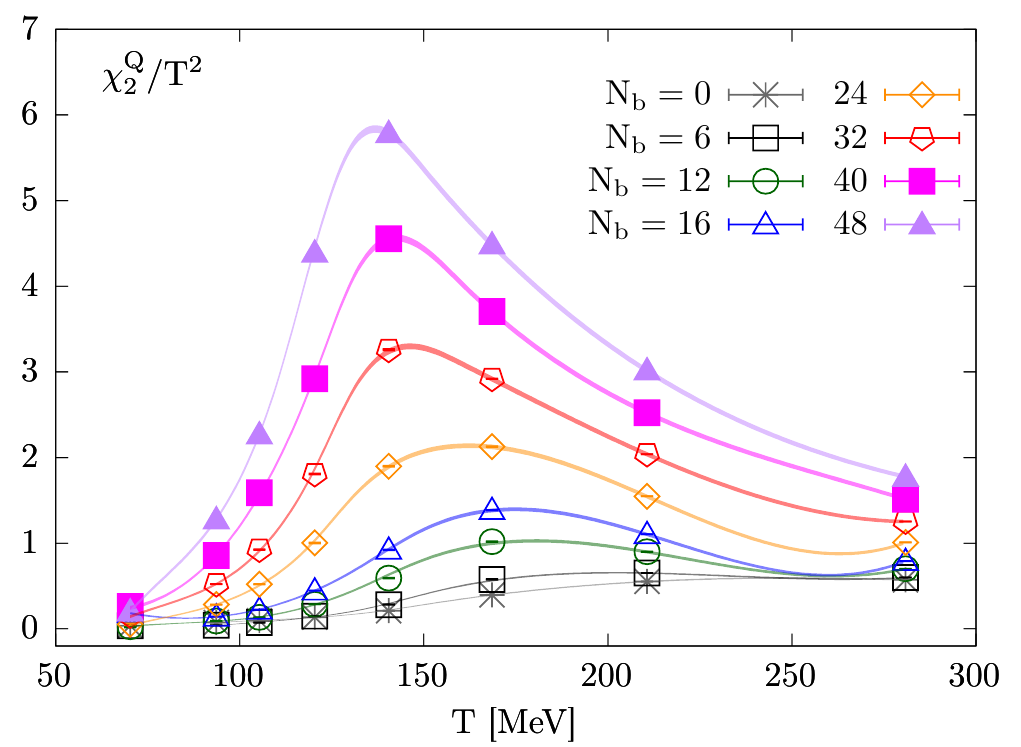}		
		\includegraphics[width=0.31\textwidth]{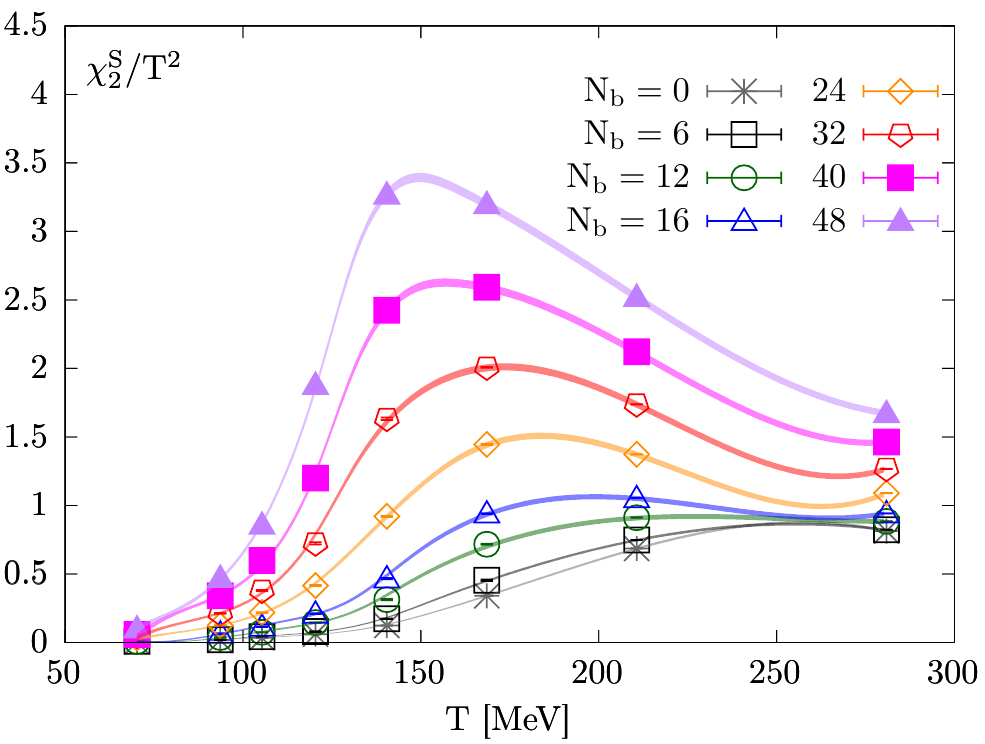}		
	\end{center}
	\caption{Temperature dependence of quadratic fluctuations of $\rm{B,Q,S}$ at various values of $N_b$. The corresponding values of $eB$ can be found in Table~\ref{tab:setup}. From left to right: $\chi_{2}^{\rm B}/T^2$, $\chi_{2}^{\rm Q}/T^2$, $\chi_{2}^{\rm S}/T^2$.
		Bands denote the spline fits to data.}	
	\label{fig:fT}
\end{figure*}

\begin{figure*}[htp!]			
	\begin{center}
		\includegraphics[width=0.31\textwidth]{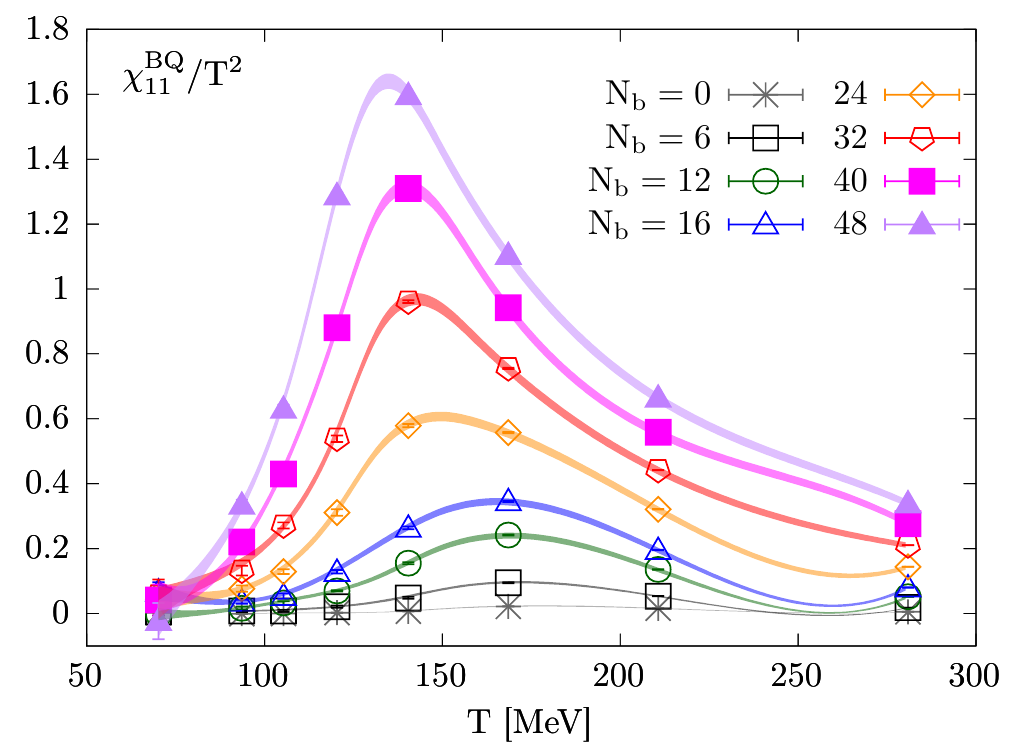}	
		\includegraphics[width=0.31\textwidth]{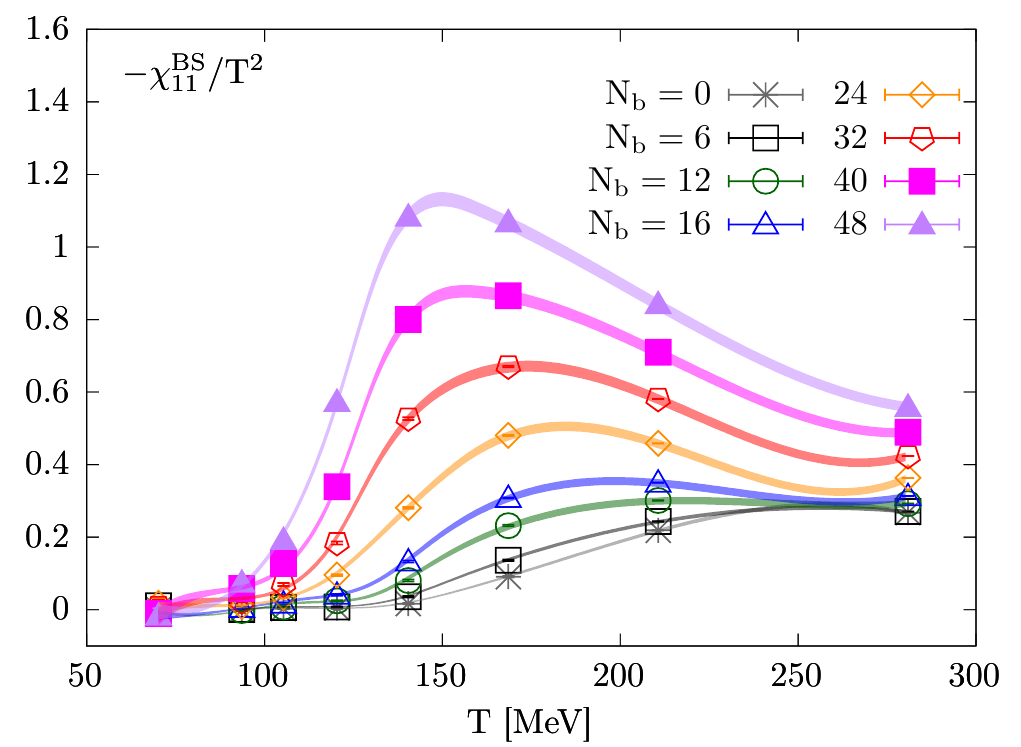}	
		\includegraphics[width=0.31\textwidth]{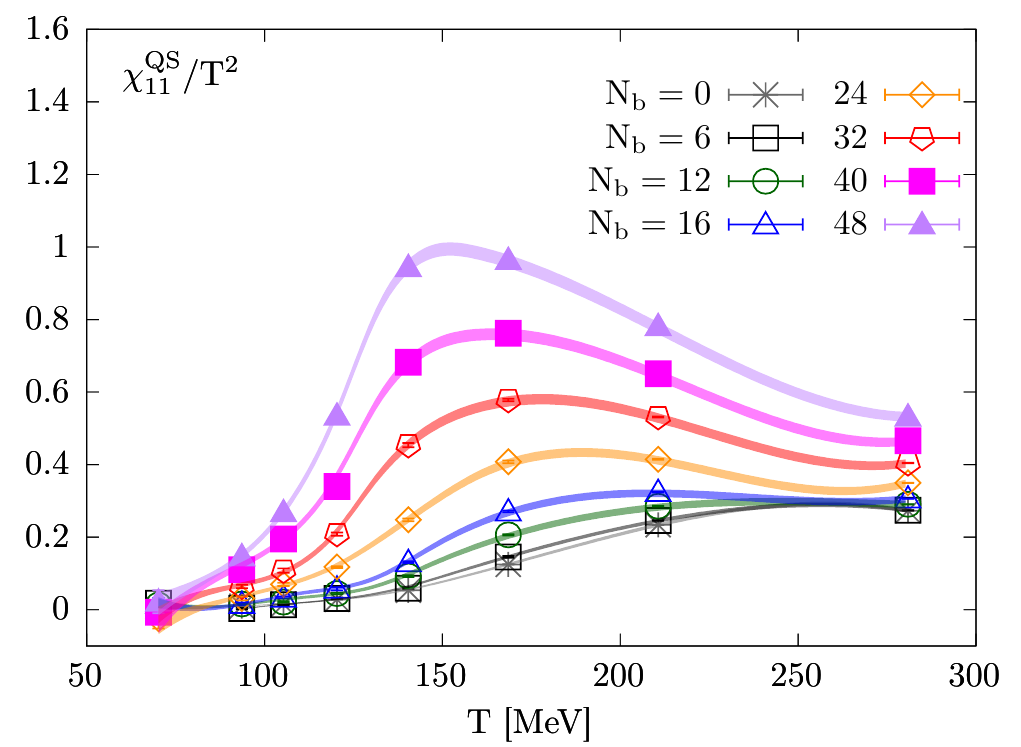} 
	\end{center}
	\caption{Same as Fig.~\ref{fig:fT} but for $\chi_{11}^{\rm BQ}/T^2$, $-\chi_{11}^{\rm BS}/T^2$ and $\chi_{11}^{\rm QS}/T^2$ from left to right.}	
	\label{fig:cT}
\end{figure*}

In our simulation $M_\pi(eB=0)\simeq220$ MeV and the resulting transition temperature at vanishing magnetic field estimated via the O(4) scaling analyses and disconnected chiral susceptibility~\cite{Ding:2019prx,Ding:2020zpp,Ding:2020rtq} is $T_{pc}(eB=0)\approx$170 MeV. 
To investigate the changes in degrees of freedom in QCD around the transition temperature, we show in Fig.~\ref{fig:fT} the temperature dependence of quadratic fluctuations of net baryon number, electric charge and strangeness, i.e. $\chi_{2}^{\rm B}$, $\chi_{2}^{\rm Q}$, $\chi_{2}^{\rm S}$ at various values of magnetic field strength $eB$. For visibility we only show results at $N_b=0$, 6, 12, 16, 24, 32, 40 and 48 which correspond to $eB/M_\pi^2(eB=0)\simeq$ 0, 6, 13, 17, 26, 34, 42 and 52, respectively. At zero magnetic field all the quadratic fluctuations of ${\rm B,~Q}$ and ${\rm S}$ increase as temperature increases, which is consistent with previous studies~\cite{Bazavov:2012jq,Bazavov:2020bjn}.  At low temperature and $eB=0$, $\chi_2^{\mathrm B}$, $\chi_2^{\mathrm Q}$ and $\chi_2^{\mathrm S}$ are dominated by the contributions from nucleon, pions and kaons, respectively.  As the magnetic field is turned on, these fluctuations start to increase faster around the transition temperature, and most strikingly they eventually develop a peak structure in strong magnetic fields. It can be clearly seen that the inflection points/peak locations of these quantities shift to lower temperatures in stronger magnetic fields. This indicates that changes in the baryon number, electric charge and strangeness carrying degrees of freedom happen at lower temperatures in stronger magnetic fields. At $eB=0$ the dissociation temperatures of nucleon, pion and kaon are relevant to the chiral crossover transition temperature determined from the chiral condensates and susceptibilities. For instance, it has been suggested that the deconfinement of strangeness happens at the chiral crossover transition temperature at $eB=0$~\cite{Bazavov:2013dta}. Thus the shifting of inflection points/peak locations of quantities shown in Fig.~\ref{fig:fT} to lower temperatures in stronger magnetic fields could be consistent with a decreasing transition temperature in larger $eB$ as determined from light quark chiral condensates and the strange quark number susceptibility~\cite{Bali:2011qj}.

On the other hand, it can also be observed from Fig.~\ref{fig:fT} that the peak height becomes higher in a stronger magnetic field. This suggests that the baryon, electric charge and strangeness carrying degree of freedom changes more rapidly across the transition in the stronger magnetic field. The higher peak and faster increasing around the transition temperature observed in the quadratic fluctuations of ${\rm B,~Q}$ and ${\rm S}$ is consistent with the finding that the strength of transition becomes larger in a stronger magnetic field~\cite{Endrodi:2015oba,Ding:2020inp}. This may signal the approach to a possible critical end point in the phase diagram in the $T$-$eB$ plane as suggested from Ref.~\cite{Endrodi:2015oba}.

We also show the quadratic correlation among ${\rm B,~Q}$ and ${\rm S}$ in Fig.~\ref{fig:cT}. $\chi_{11}^{\rm BQ}$ , which denotes the correlation between baryon number and electric charge, is dominated by the contribution from protons at low temperature and goes to zero in the high-temperature limit with vanishing quark masses. It thus naturally develops a peak structure already at zero magnetic field~\cite{Bazavov:2011nk}, which can also be observed in our current study. At nonzero magnetic fields, the peak structure in $\chi_{11}^{\rm BQ}$ becomes more striking and the peak location also shifts to lower temperatures in the stronger magnetic field. $-\chi_{11}^{\rm BS}$ and $\chi_{11}^{\rm QS}$, as shown in the middle and right panel of Fig.~\ref{fig:cT}, respectively, possess similar features as seen in $\chi_2^{\rm B,Q,S}$.

\subsection{Isospin symmetry breaking effects at nonzero magnetic fields}
\label{sec:resiso}
\begin{figure}[htp]			
	\begin{center}
		\includegraphics[width=0.33\textwidth]{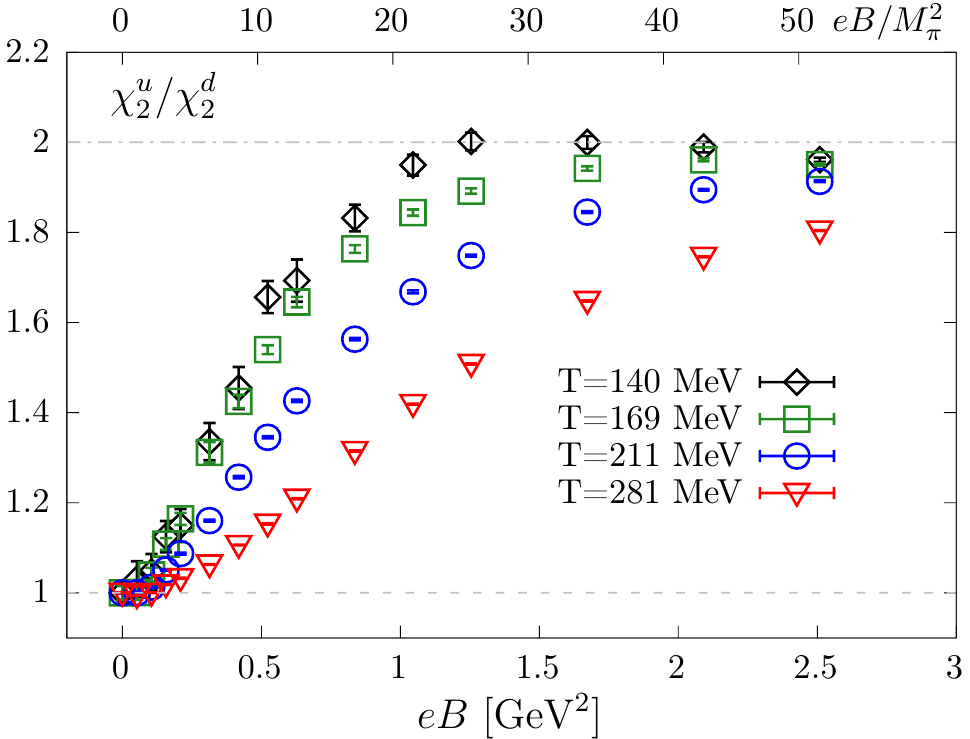}\\
		\includegraphics[width=0.33\textwidth]{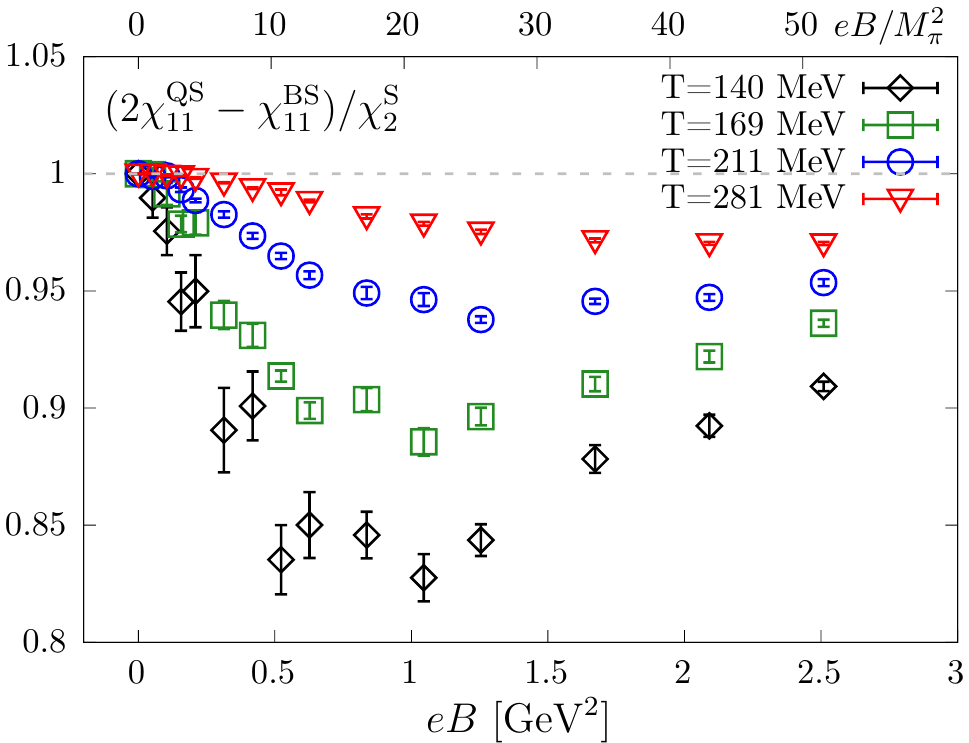}\\
		\includegraphics[width=0.33\textwidth]{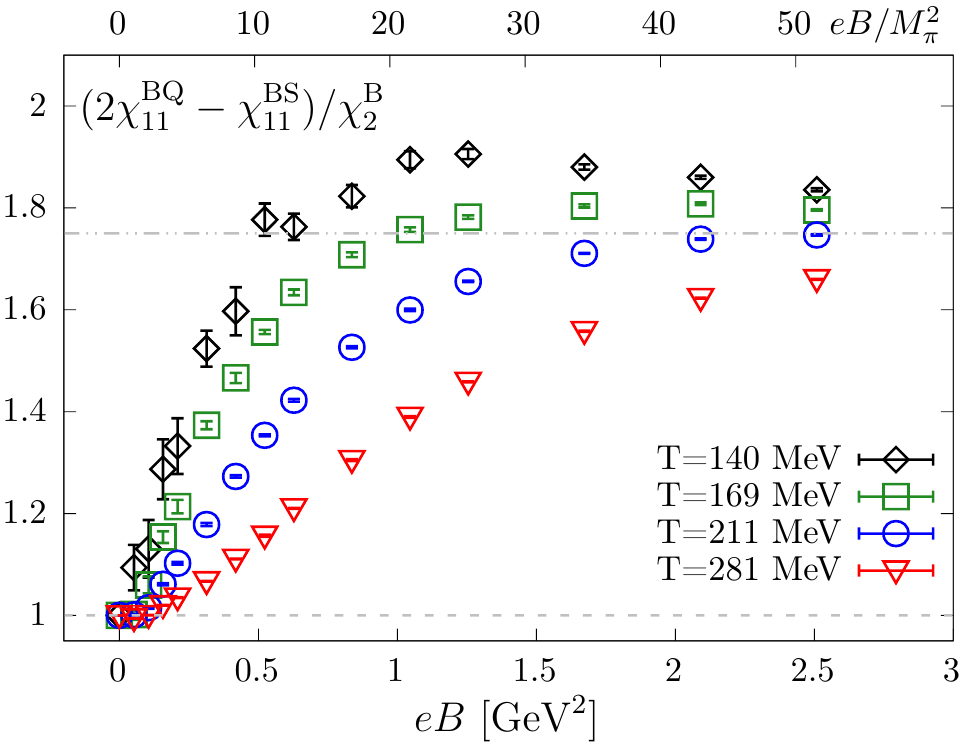}
	\end{center}
	\caption{Isospin symmetry breaking effects manifested in $\chi_2^u/\chi_2^d$ (top), $(2\chi_{11}^{\rm QS}-\chi_{11}^{\rm BS})/\chi_2^{\rm S}$ (middle) and $(2\chi_{11}^{\rm BQ}-\chi_{11}^{\rm BS})/\chi_2^{\rm BQ}$ (bottom). The dash-dotted lines in all the plots represent the ideal gas limits and dashed lines denote results in the isospin symmetric case.}	 
	\label{fig:iso}
\end{figure}

In our lattice simulation, the up and down quark masses are degenerate at $eB=0$. Since up and down quarks have different electric charge, the isospin symmetry is obviously broken once the magnetic field is turned on. As seen from the top panel of Fig.~\ref{fig:iso} the ratio of up to down quark number susceptibility, $\chi_2^u/\chi_2^d$, is unity at all temperatures at $eB=0$, and becomes larger than 1 at $eB\neq0$. As in the ideal gas limit with $\sqrt{eB}/T\rightarrow\infty$ $\chi_2^u/\chi_2^d$ equals to 2, it is expected that $\chi_2^u/\chi_2^d$ increases from 1 towards 2 as $eB$ grows. Results shown in the top panel of Fig.~\ref{fig:iso} are consistent with this expectation. It is also interesting to see that $\chi_2^u/\chi_2^d$ increases faster at lower temperatures. This suggests that the isospin symmetry is broken more seriously at lower temperatures at a fixed value of $eB$.

We further investigate the isospin symmetry breaking effects at the level of B, Q and S. At $eB=0$ due to the isospin symmetry of up and down quarks, the six quadratic fluctuations and correlations of ${\rm B,Q}$ and ${\rm S}$ are not independent and constrained by the following two relations as $\chi_{11}^{us}=\chi_{11}^{ds}$
\begin{align}
\label{eq:iso1}
    2\chi_{11}^{\rm QS}-\chi_{11}^{\rm BS} &=\chi_2^{\rm S}, \\
    2\chi_{11}^{\rm BQ}-\chi_{11}^{\rm BS} &=\chi_2^{\rm B}.
    \label{eq:iso2}
\end{align}
As a consequence of Eq.~\ref{eq:free2}, Eq.~\ref{eq:iso1} also holds true in the ideal gas limit with $eB\neq0$. $(2\chi_{11}^{\rm QS}-\chi_{11}^{\rm BS})/\chi_2^{\rm S}$ thus equals to unity at all temperatures with $eB=0$ and at high temperatures with $eB\neq0$.
This is exactly what can be seen from the middle panel of Fig.~\ref{fig:iso}. At $eB=0$ the ratio $(2\chi_{11}^{\rm QS}-\chi_{11}^{\rm BS})/\chi_2^{\rm S}$ is unity at all four temperatures and then starts to decrease as the magnetic field grows and finally has to approach to unity after a turning point. Similarly as observed from the top panel of Fig.~\ref{fig:iso} the ratio changes more rapidly as a function of $eB$ at lower temperatures.

On the other hand Eq.~\ref{eq:iso2} holds at any temperature with $eB=0$, however, it does not hold true any more with $eB\neq0$ in the ideal gas limit. In the ideal gas limit, where the correlations among $u$, $d$ and $s$ vanish, $2\chi_{11}^{\rm BQ}-\chi_{11}^{\rm BS}$ equals to $(4\chi_2^u-\chi_2^d)/9$ while $\chi_2^{\rm B}=(\chi_2^u+2\chi_2^d)/9$. In the ideal gas limit with $\sqrt{eB}/T\rightarrow \infty$  the ratio of $(2\chi_{11}^{\rm BQ}-\chi_{11}^{\rm BS})/\chi_2^{\rm B}$ thus approaches to 7/4 as $\chi_2^u=2\chi_2^d$ (cf. Table~\ref{tab:largeBfree}). Values of $(2\chi_{11}^{\rm BQ}-\chi_{11}^{\rm BS})/\chi_2^{\rm B}$ at $eB=0$ and in the high-temperature limit with $\sqrt{eB}/T\rightarrow\infty$ thus suggest a monotonous increasing behavior of the ratio as a function of $eB$. This, however, is only the case for two highest temperatures of 211 and 281 MeV, as seen from the bottom panel of Fig.~\ref{fig:iso}. For lower temperatures, i.e. 169 and 140 MeV, the free limit is approached from above and the ratio thus develops a weak non-monotonous behavior as a function of $eB$. We remark that isospin symmetry breaking effects are mostly manifested at lower temperatures in all three quantities shown in Fig.~\ref{fig:iso}.

In the heavy-ion collisions the strength of the magnetic field produced in the initial collisions is about ${0-0.6}$ GeV$^2$~\cite{skokov2009estimate}. This corresponds to $0-12 M_\pi^2(eB=0)$ with $M_\pi(eB=0)\simeq220$ MeV in our lattice setup. To probe isospin symmetry breaking effects experimentally, one in principle could look at $\chi_2^u/\chi_2^d$ expressed in the terms of quadratic fluctuations and correlations of ${\rm B,~Q}$ and ${\rm S}$~(cf.$\linebreak$~Eq.~\ref{eq:chiudBQS}). However, precise measurements of right hands of Eq.~\ref{eq:chiudBQS} in heavy-ion collision experiments could be difficult. As can be observed in Fig.~\ref{fig:iso} the deviation from the isospin symmetric case is even larger in $(2\chi_{11}^{\rm BQ}-\chi_{11}^{\rm BS})/\chi_2^{\rm B}$ than in $\chi_2^u/\chi_2^d$. For instance at $eB\simeq0.5$ GeV$^2$, the former deviation is about 50\% while the latter is about 80\%. Thus this could render $(2\chi_{11}^{\rm BQ}-\chi_{11}^{\rm BS})/\chi_2^{\rm B}$ a useful probe for the isospin symmetry breaking~\footnote{One can also construct quantities without $\chi_2^{\rm S}$ to reflect the isospin symmetry breaking, e.g. $(\chi_2^{\rm Q}- 2\chi_{11}^{\rm BQ})/(\chi_2^{\rm Q}+\chi_{11}^{\rm BQ})$, and $0.5(2\chi_2^{\rm B}- \chi_{11}^{\rm BQ})/(\chi_2^{\rm B}+\chi_{11}^{\rm BQ})$. Both of these two quantities approach to $\chi_2^d/\chi_2^u$ in the high-temperature limit.}.

\subsection{Comparisons to Hadron Resonance Gas model \& high-temperature free limit}
\label{sec:resHRGfreePH}

At low temperatures and zero magnetic fields QCD thermodynamics can be well described by the hadron resonance gas model~\cite{Ding:2015ona}. In the nonzero magnetic fields, the situation becomes complex as the hadron spectra are modified by the magnetic field. It has been found that energies of charged particles, e.g. $\pi^{+,-} (K^{+,-})$ obey the lowest Landau-level (cf. Eq.~\ref{eq:Elevel}) only at $eB\lesssim 0.31$ GeV$^2$ and then turn out to deviate from the the lowest Landau-level and finally decrease at $eB\gtrsim$ 0.5 GeV$^2$, while those of neutral particles, e.g. neutral pion decreases as $eB$ grows in full QCD~\cite{Ding:2020hxw}. Since the $eB$-dependence of neutral particles' masses (besides $\pi^0$, $K^0$, neutron, $\Sigma^0$ and $\Xi^0$~\cite{Martinelli:1982cb,Chang:2015qxa,Parreno:2016fwu,Ding:2020hxw,Endrodi:2019whh}) have not been studied yet in lattice QCD computations, we thus focus on the fluctuations and correlations involving electric charge Q, $\chi_{11}^{\rm BQ}$, $\chi_2^{\rm Q}$ and $\chi_{11}^{\rm QS}$ which receive no contributions from neutral particles. On the other hand, the energy of charged hadron obeys the lowest Landau-level as shown in Eq.~\ref{eq:Elevel} at $eB\lesssim0.31$ GeV$^2$, in which we have 4 values of $eB$ at each temperature. We thus focus on the comparison with HRG results in the case of  $eB\lesssim 0.31$ GeV$^2$. 

\begin{figure*}[htp!]			
	\begin{center}
		\includegraphics[width=0.32\textwidth]{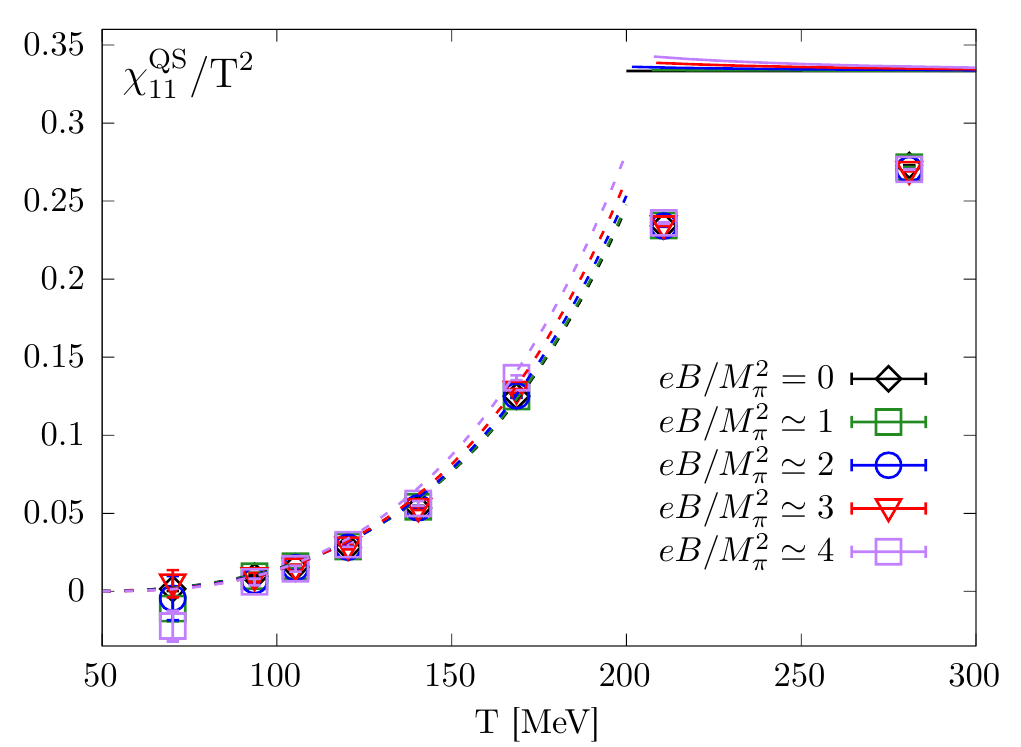} 
		\includegraphics[width=0.32\textwidth]{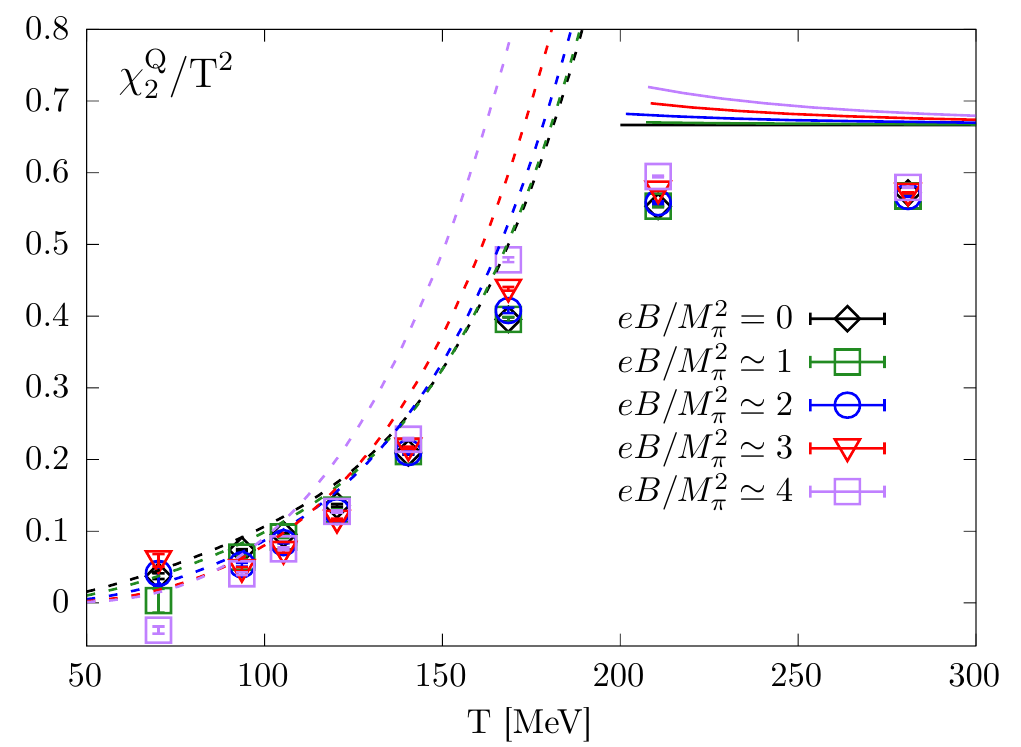} 
		\includegraphics[width=0.32\textwidth]{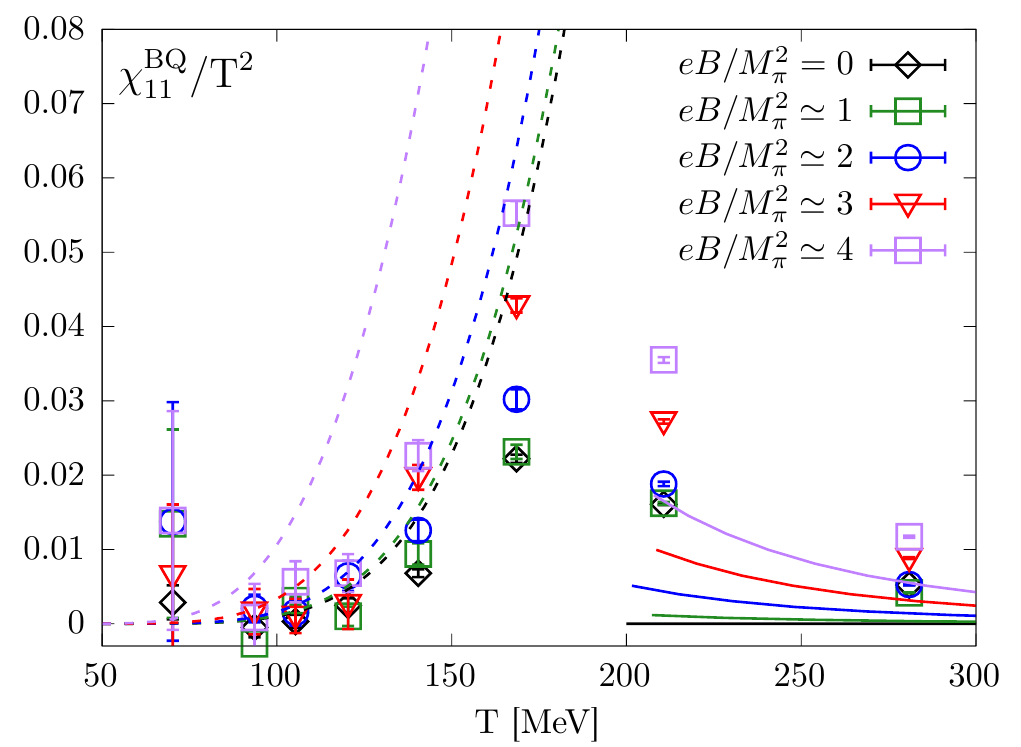} 
	\end{center}
	\caption{Comparisons of $\chi_{11}^{\rm QS}/T^2$ (left), $\chi_{2}^{\rm Q}/T^2$ (middle) and $\chi_{11}^{\rm BQ}/T^2$ (right) with results obtained from the HRG model and ideal gas limit. The dashed lines having the same colors as the lattice data denote corresponding HRG results using hadron spectrum obtained from PDG~\cite{Tanabashi:2018oca}, while the solid lines represent the corresponding free limits at each value of $eB$. For visibility the solid lines are plotted starting from different temperature values.}	 
	\label{fig:compHRG}
\end{figure*}
\begin{figure*}[h!]			
	\begin{center}
		\includegraphics[width=0.32\textwidth]{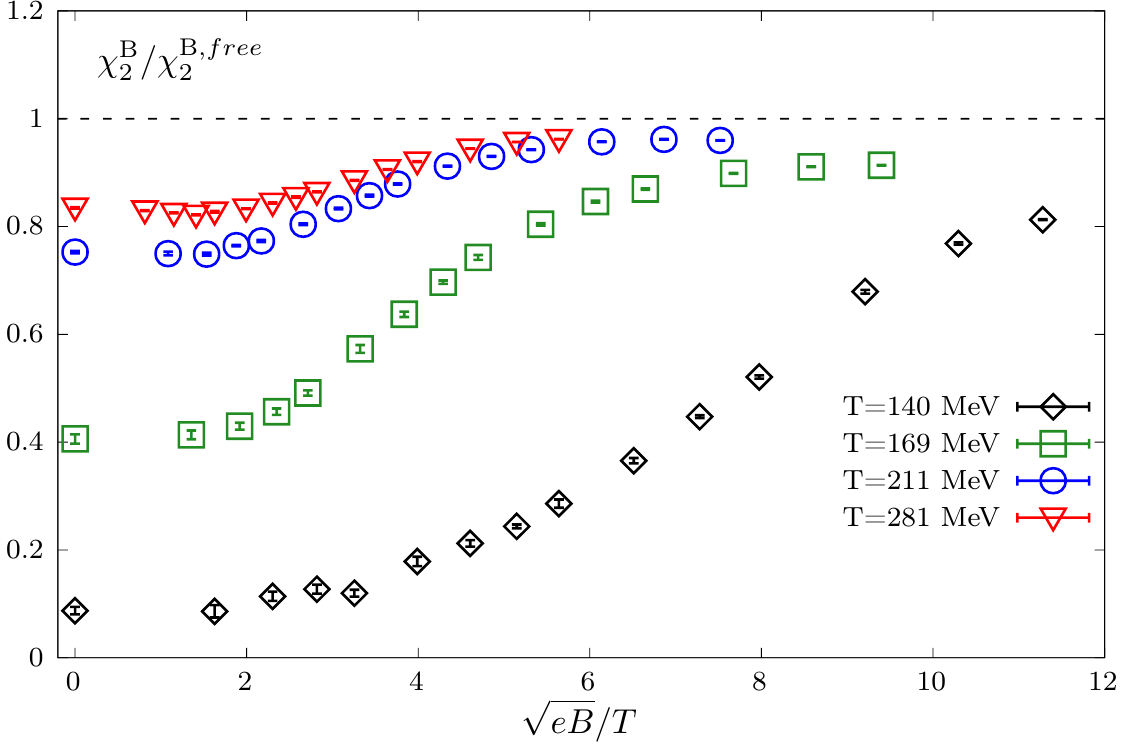}		\includegraphics[width=0.32\textwidth]{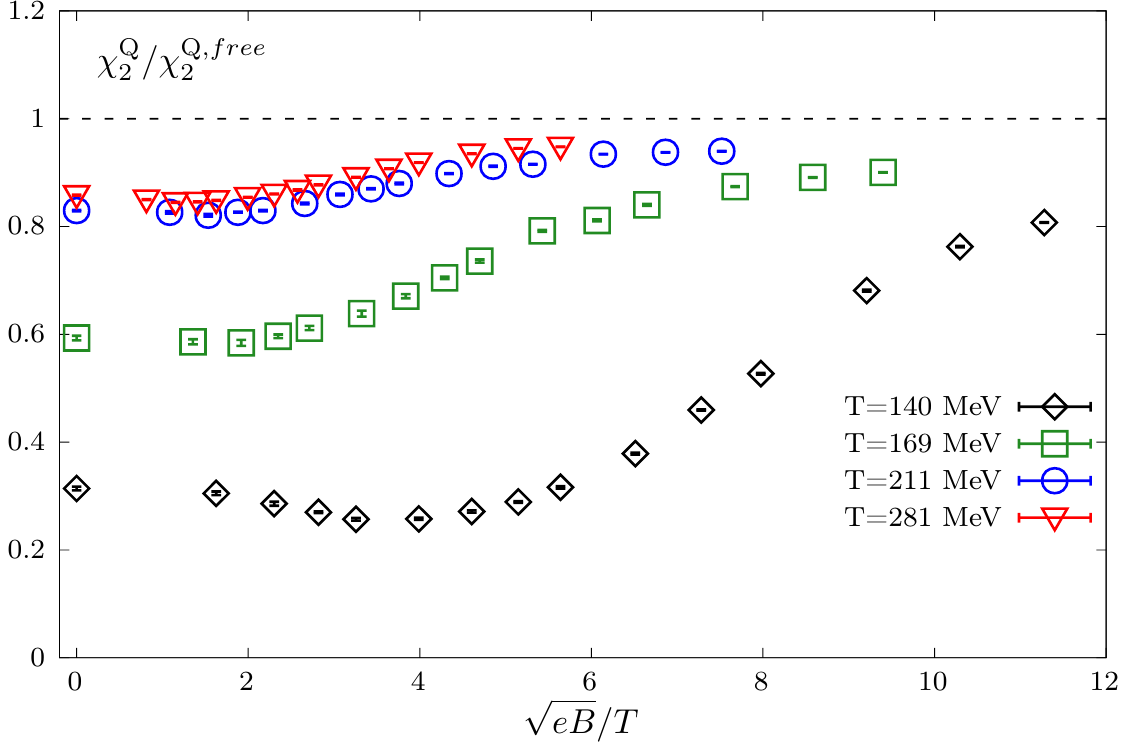}		\includegraphics[width=0.32\textwidth]{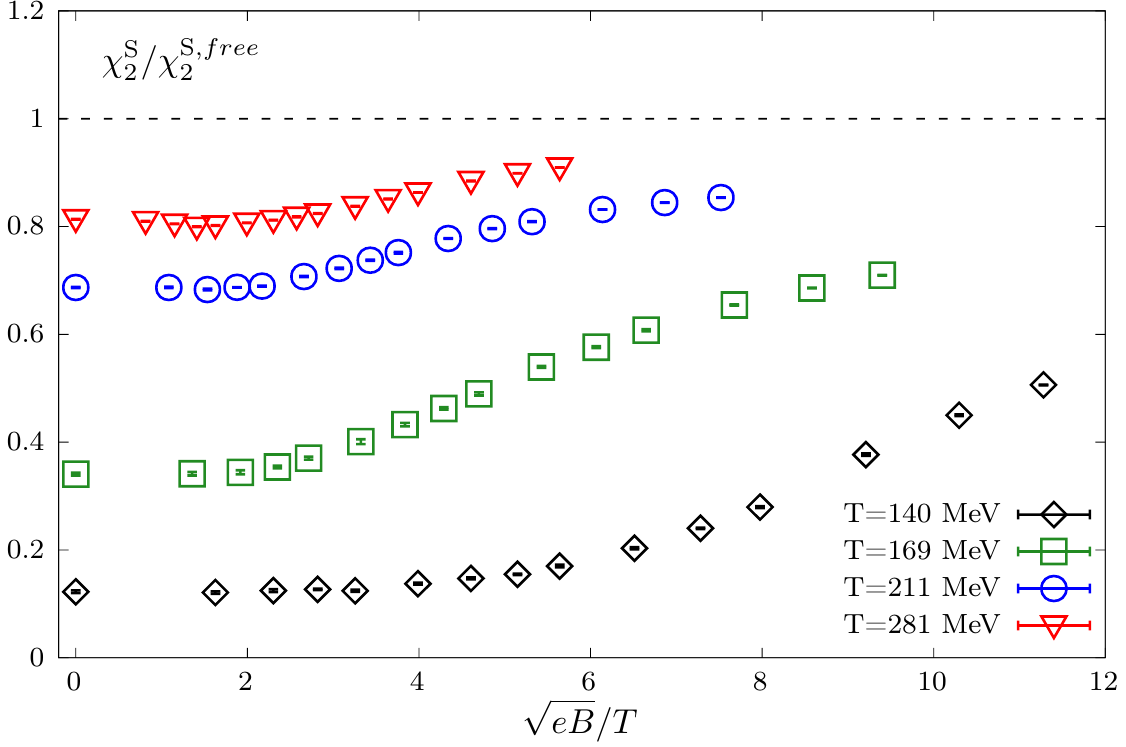}
		\includegraphics[width=0.32\textwidth]{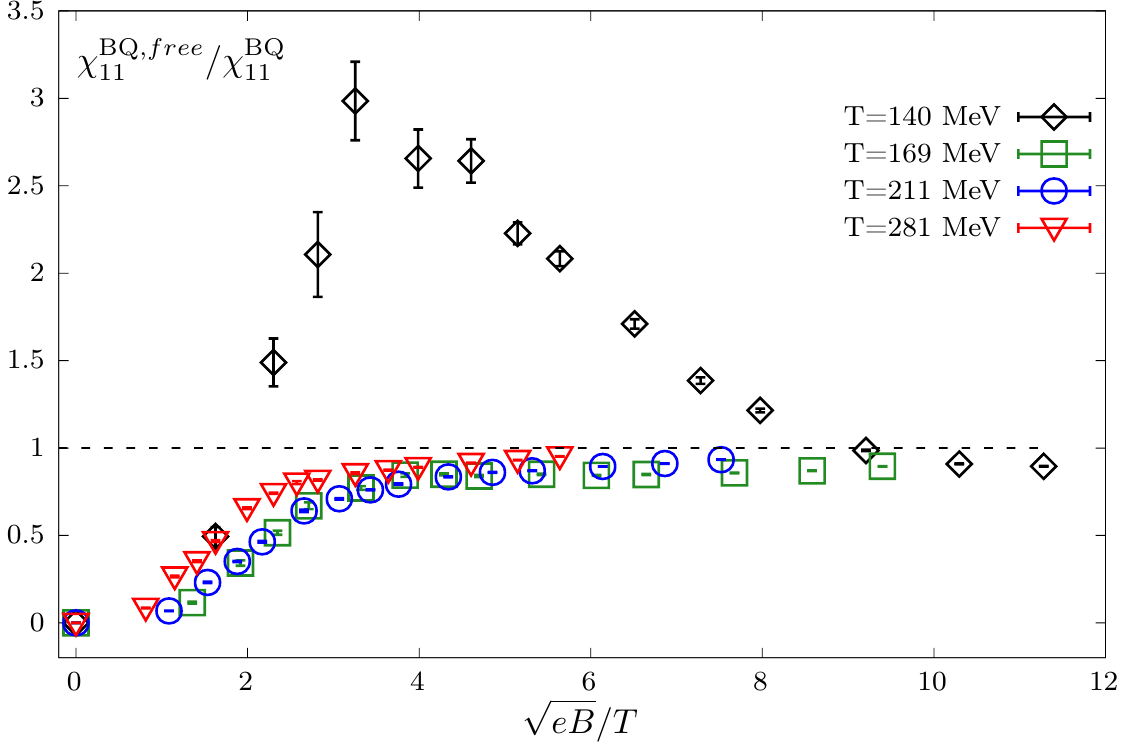}
		\includegraphics[width=0.32\textwidth]{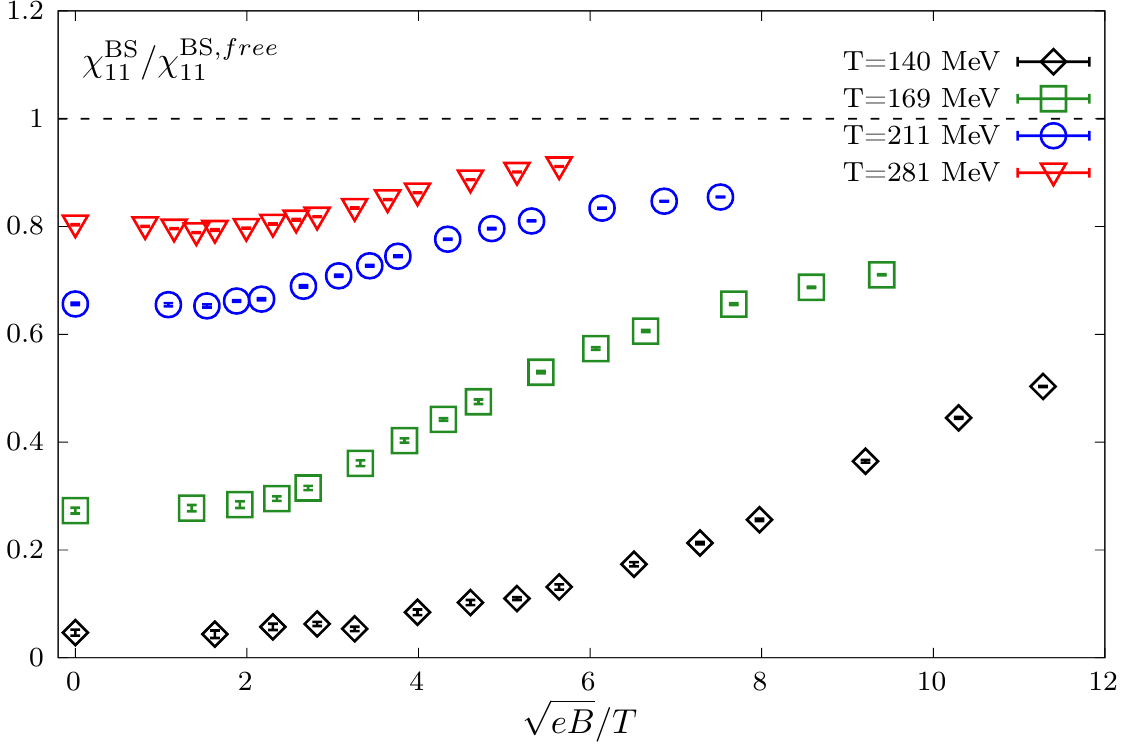}	
		\includegraphics[width=0.32\textwidth]{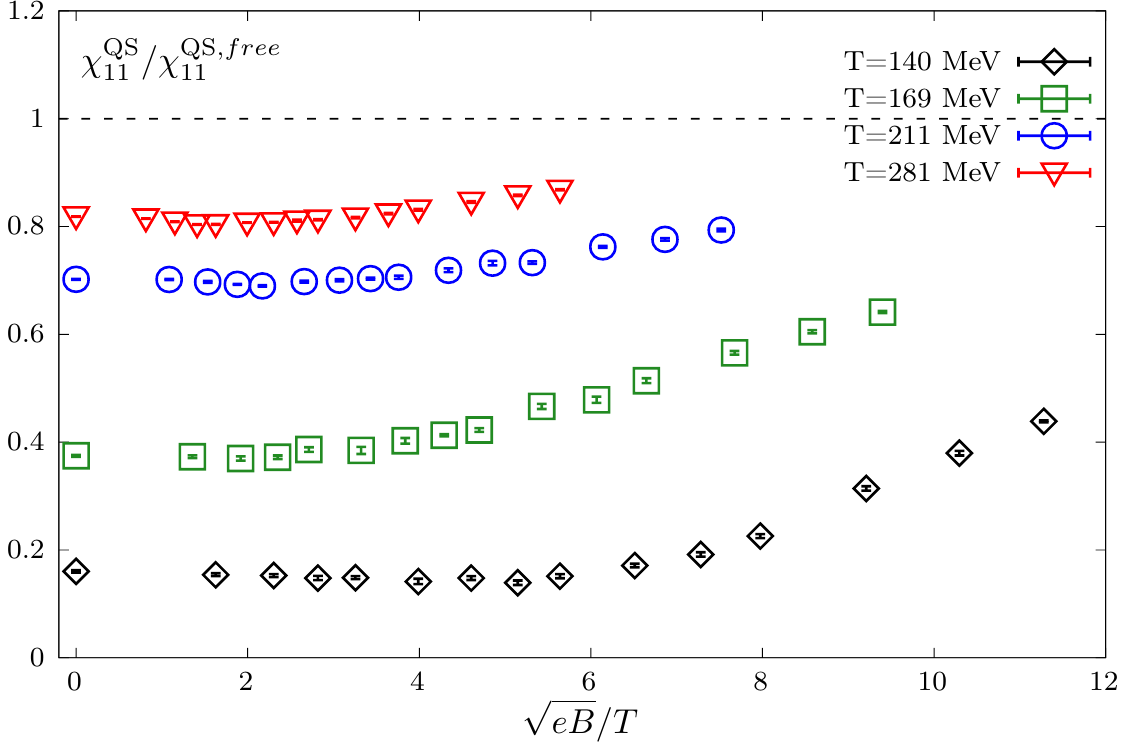}\\	
	\end{center}
	\caption{Ratios to corresponding ideal gas limits as a function of $\sqrt{eB}/T$. Top: $\chi_2^\mathrm{B}/\chi_2^{\mathrm{B},free}$, $\chi_2^\mathrm{Q}/\chi_2^{\mathrm{Q},free}$ and $\chi_2^\mathrm{S}/\chi_2^{\mathrm{S},free}$ from left to right. Bottom: $\chi_{11}^{\mathrm{BQ},free}/\chi_{11}^\mathrm{BQ}$, $\chi_{11}^\mathrm{BS}/\chi_{11}^{\mathrm{BS},free}$ and $\chi_{11}^\mathrm{QS}/\chi_{11}^{\mathrm{QS},free}$ from left to right.}	 
	\label{fig:free}
\end{figure*}

In Fig.~\ref{fig:compHRG} we show lattice data of  $\chi_{11}^{\rm QS}/T^2$(left),  $\chi_{2}^{\rm Q}/T^2$ (middle) and $\chi_{11}^{\rm BQ}/T^2$ (right) as functions of temperature at various values of $eB$ with $N_b=0,$ 1, 2, 3 and 4 corresponding to $eB/M_\pi^2=0$ and $eB/M_\pi^2\simeq$ 1, 2, 3 and 4, respectively. Also shown are the results obtained from HRG (cf. Eq.~\ref{eq:HRGresults})~\footnote{Here in the HRG calculations we adopt the PDG hadron spectrum except that at $eB=0$ masses of pion, kaon and $\rho$ determined in our current lattice setup are used instead of those listed in PDG.} denoted as dashed lines, and from ideal gas limit (cf. Eqs.~\ref{eq:free11QS},~\ref{eq:free2Q} and~\ref{eq:free11BQ}) denoted as solid lines.  It can be seen from the left panel of Fig.~\ref{fig:compHRG} that $\chi_{11}^{\rm QS}$, which is dominated by charged kaons at low temperatures, is almost $eB$ independent within the current $eB$ window at $T\lesssim281$ MeV\footnote{The $eB$-dependence seen at $T\simeq70$ MeV could be due to the statistics-hungry nature of the observables at low temperature and insufficient statistics we have in the simulation, similar in the cases of $\chi_{2}^{\rm Q}$ and $\chi_{11}^{\rm BQ}$.}. On the other hand, the HRG results, which grow exponentially from zero, also show mild $eB$-dependence and give a good description of the lattice data of $\chi_{11}^{\rm QS}$ at $T\lesssim 169$ MeV. 
\begin{figure*}[h!]			
	\begin{center}
		\includegraphics[width=0.32\textwidth]{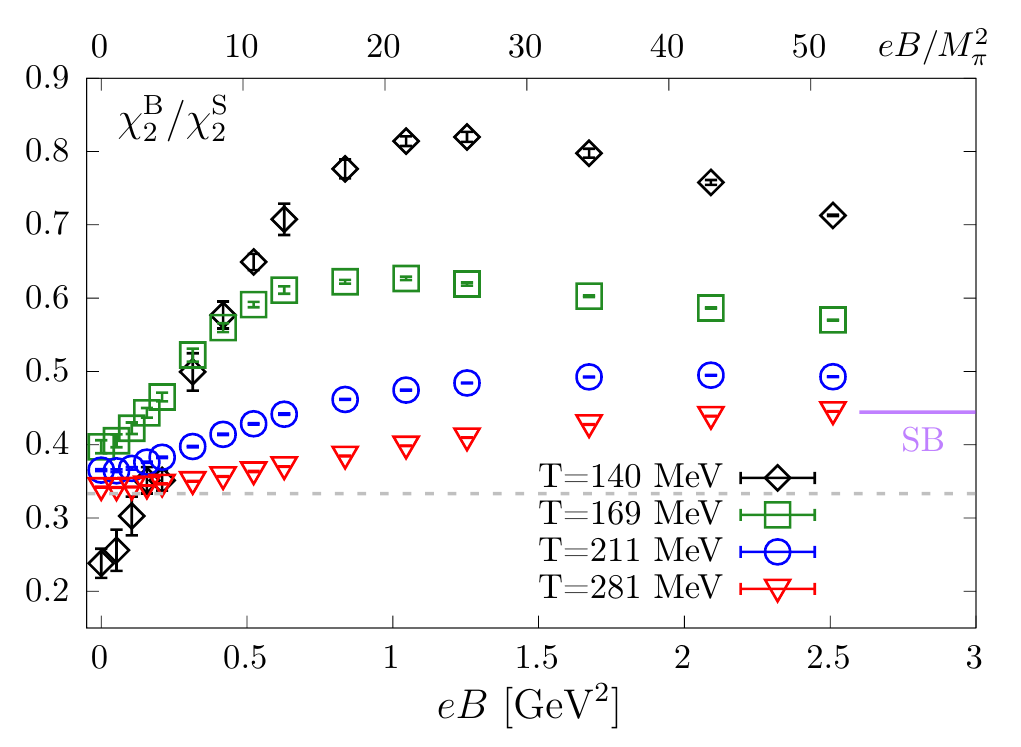}	
		\includegraphics[width=0.32\textwidth]{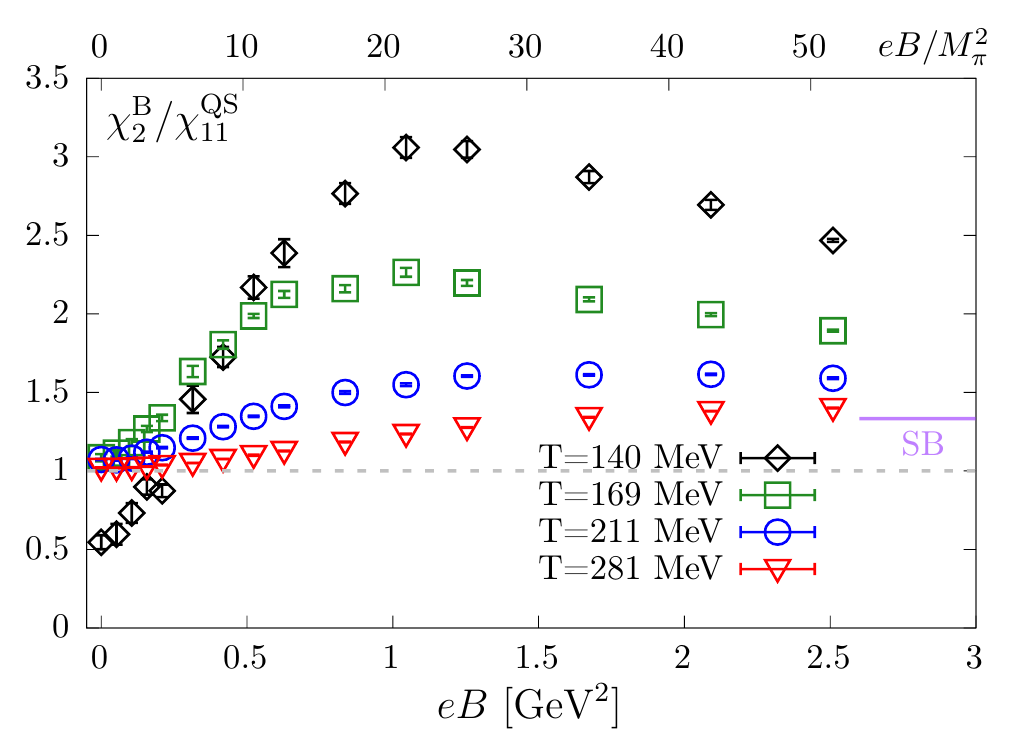}	
		\includegraphics[width=0.32\textwidth]{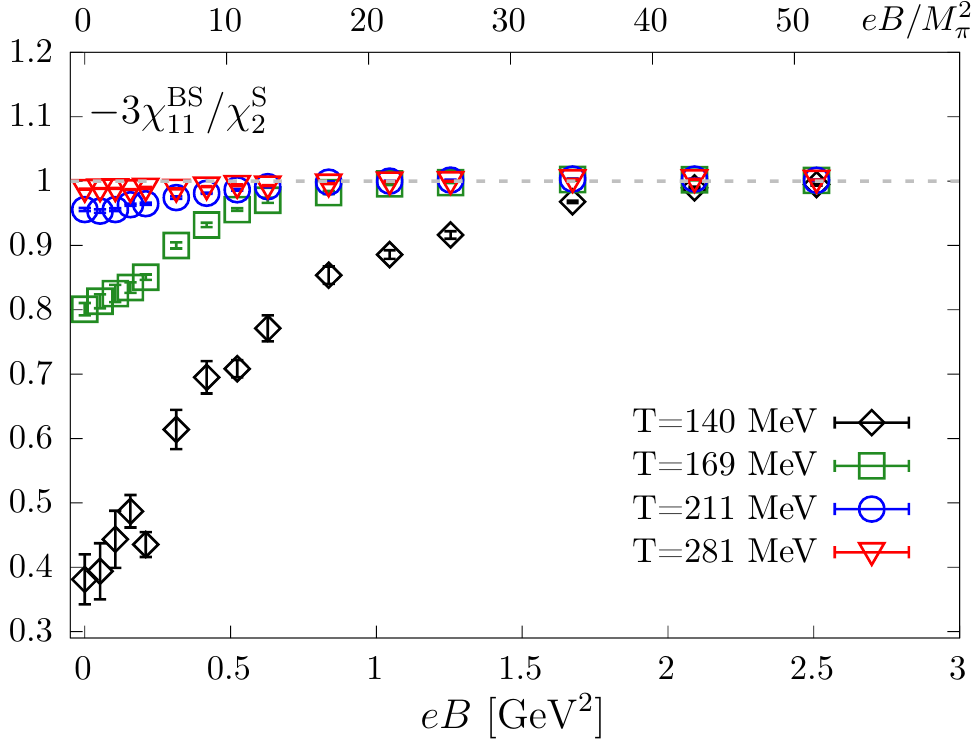}
	\end{center}
	\caption{$\chi_{2}^{\rm B}/\chi_{2}^{\rm S}$ (left), $\chi_{2}^{\rm B}/\chi_{11}^{\rm QS}$ (middle) and $-3\chi_{11}^{BS}/\chi_2^S$(right) as a function of $eB$. The dashed lines denote the ideal gas limits at $eB=0$ while the solid lines denote the ideal gas limits with $\sqrt{eB}/T\rightarrow\infty$.}	 
	\label{fig:ratio_ph}
\end{figure*}
The lattice data of $\chi_2^{\rm Q}$ has slightly larger $eB$-dependence compared to $\chi_{11}^{\rm QS}$ only at $T\simeq169$ MeV. This might be understood that the pion masses are more affected compared to kaons by the magnetic field~\cite{Ding:2020hxw}. As at $eB=0$ $\chi_2^{\rm Q}$ is dominated by charged pions at low temperatures and the pion spectrum is strongly affected by the taste symmetry violation in the staggered formalism, here in the HRG computation of $\chi_2^{\rm Q}$ we adopt the corresponding root-mean-squared pion mass instead of the Goldstone pion mass.  It can be seen from the middle panel of Fig.~\ref{fig:compHRG} that the HRG results start to have considerable $eB$-dependences already at $T\gtrsim140$ MeV, and they can only describe the lattice data reasonably well at $T\lesssim140$ MeV. 

In the right panel of Fig.~\ref{fig:compHRG} it can be seen that the lattice data of $\chi_{11}^{\rm BQ}$ possess the largest $eB$-dependence among the three observables considered here, i.e. significant effects induced by $eB$ are clearly shown already at $T\gtrsim140$ MeV. This might be due to the complex $eB$ dependence of charged baryons~\cite{Endrodi:2019whh}. The HRG results for $\chi_{11}^{\rm BQ}$, similar as those for $\chi_{11}^{Q}$ and $\chi_{11}^{QS}$, increase exponentially in temperature from zero to higher values and start to be nonzero at lower temperatures as $eB$ grows. Thus the description of HRG to both $\chi_{11}^{\rm BQ}$ and $\chi_{2}^{\rm Q}$ breaks down at lower temperatures as $eB$ grows. This is consistent with the fact that the transition temperature becomes lower with larger $eB$ as HRG is supposed to describe the lattice data only in the low-temperature phase of QCD.

In Fig.~\ref{fig:compHRG} $\chi_{11}^{\rm QS}/T^2$, $\chi_{2}^{\rm Q}/T^2$ and $\chi_{11}^{\rm BQ}/T^2$ are observed to approach to the ideal gas limit as temperature and $eB$ become larger.
As discussed in Section~\ref{sec:subfree} the quadratic fluctuations and correlations of B, Q and S scale with $\sqrt{eB}/T$ in the high-temperature limit, we show in Fig.~\ref{fig:free} the five quadratic fluctuations and correlations of B, Q and S divided by their corresponding values in the free limit as function of $\sqrt{eB}/T$(cf. Eqs.~\ref{eq:freeB}-\ref{eq:freeBQS}), and for  $\chi_{11}^{\rm BQ}$, we rather show $\chi_{11}^{{\rm BQ},free}/\chi_{11}^{{\rm BQ}}$ as $\chi_{11}^{\rm BQ}=0$ in the ideal gas limit at $eB=0$. One can clearly see that all the six ratios approach to 1 as $\sqrt{eB}/T$ grows at all four different temperatures, and the ratios increase faster at lower temperatures. The former observation can be understood as the temperature divided by $T_{pc}(eB)$ becomes larger in the stronger magnetic fields since the transition temperature $T_{pc}(eB)$ reduces as $eB$ grows. The latter observation could be mainly because at a lower temperature, e.g. $T=$ 140 MeV, the degree of freedom in the system changes dramatically from confined hadron phase to deconfined quark-gluon plasma phase, while at the high temperature, e.g. $T \gtrsim$ 211 MeV, the system is already in the deconfined quark-gluon plasma phase at $eB=0$, and is just pushed deeper into the deconfined phase with increasing $eB$. The other interesting observation is that all the fluctuations and correlations of B, Q and S at all temperatures except $\chi_{11}^{\rm BQ}$ approach to the free limit from below. $\chi_{11}^{\rm BQ}$, on the other hand, approach to the free limit from above at $T=$169, 211 and 281 MeV, and firstly approach its free limit from above and then from below at $T=140$ MeV. This might be explained as vanishing B-Q correlation in the high-temperature limit at $eB=0$ and the complex baryon spectrum in the magnetic field.

In Fig.~\ref{fig:ratio_ph} we show ratios of $\chi_{2}^{\rm B}/\chi_{2}^{\rm S}$ (left), $\chi_{2}^{\rm B}/\chi_{11}^{\rm QS}$ (middle) and $-3\chi_{11}^{BS}/\chi_2^S$(right) as function of $eB$ in the phenomenologically interesting temperature region $T\gtrsim$ 140 MeV. $\chi_{2}^{\rm B}/\chi_{2}^{\rm S}$ equals to 1/3 in the ideal gas limit at $eB=0$, while it increases to 4/9 in the ideal gas limit with $\sqrt{eB}/T\rightarrow\infty$. At $eB=0$ $\chi_{2}^{\rm B}/\chi_{2}^{\rm S}$ approaches the ideal gas limit from above in the current temperature window. As the magnetic field is turned on, $\chi_{2}^{\rm B}/\chi_{2}^{\rm S}$ at $T=281$ MeV increases slowly as $eB$ grows and tends to approach the ideal gas limit with $\sqrt{eB}/T\rightarrow\infty$ from the above as well. As temperature becomes lower $\chi_{2}^{\rm B}/\chi_{2}^{\rm S}$ increases faster in $eB$ and develops a non-monotonous behavior in $eB$ at the two lowest temperatures. And at a lower temperature $\chi_{2}^{\rm B}/\chi_{2}^{\rm S}$ also starts to decrease at  a smaller value of $eB$. Similar features can also be observed in $\chi_{2}^{\rm B}/\chi_{11}^{\rm QS}$.

The ratio of baryon-strangeness correlation to $\linebreak$strangeness fluctuation, $-3\chi_{11}^{\rm BS}/\chi_{2}^{\rm S}$, as shown in the right panel of Fig.~\ref{fig:ratio_ph},is also of interest. The free limit of  $-3\chi_{11}^{\rm BS}/\chi_{2}^{\rm S}$, no matter whether the magnetic field is present or not, is always 1. At $T=$ 281 MeV, $-3\chi_{11}^{\rm BS}/\chi_{2}^{\rm S}$ already equals to 1 at $eB=0$, and the presence of magnetic field thus does not bring any change to this quantity. At lower temperatures, i.e. $T<281$ MeV, where  $-3\chi_{11}^{\rm BS}/\chi_{2}^{\rm S}<1$ at $eB$=0, the presence of magnetic field thus brings the ratio up towards to 1. Similar as learned before from Fig.~\ref{fig:free}, $\chi_{2}^{\rm B}/\chi_{2}^{\rm S}$ and $\chi_{2}^{\rm B}/\chi_{11}^{\rm QS}$, $-3\chi_{11}^{\rm BS}/\chi_{2}^{\rm S}$ has the most significant $eB-$dependence at the lowest temperature. This suggests that the magnetic field fosters the transition, which is consistent with the fact that the transition temperature reduces as $eB$ grows. In particular, even at $T=$140 MeV $-3\chi_{11}^{\rm BS}/\chi_{2}^{\rm S}$ can be induced to its free limit with $eB\gtrsim2$ GeV$^2$, while $\chi_{2}^{\rm B}/\chi_{2}^{\rm S}$ and $\chi_{2}^{\rm B}/\chi_{11}^{\rm BS}$ are about 60\% and 80\% away from the free limit, respectively. 

As discussed in previous subsection, $eB$ produced in heavy-ion collision experiments can reach up to $\sim$12$M_\pi^2$. At $eB\sim0.5$ GeV$^2$ ($N_b=10$) for instance, the ratios of $\chi_2^{\rm B}/\chi_2^{\rm S}$ divided by its value at $eB=0$ are about 2.7 and 1.5 at $T=140$ and $169$ MeV, respectively.  For $-3\chi_{11}^{\rm BS}/\chi_{2}^{\rm S}$ the ratios are about 1.8 at $T=140$ MeV and 1.2 at $T=169$ MeV. The change of $\chi_2^{\rm B}/\chi_{11}^{\rm QS}$ at $eB\sim0.5$ GeV$^2$ as compared to the case of $eB=0$ is most significant, i.e. the ratios are about 4 at $T=$140 MeV and 1.8 at $T=$169 MeV.

\section{Discussion and conclusion}
\label{sec:summary}

In this paper we present the first results of quadratic fluctuations and correlations of net baryon number, electric charge, and strangeness from lattice QCD simulations in nonzero magnetic fields. The lattice QCD simulations are performed using the HISQ fermions with a tree-level improved Symanzik gauge action with $M_\pi(eB=0)\simeq220$ MeV. The fixed scale approach with lattice spacing $a\simeq0.117$ fm is adopted such that $eB$ only varies with the magnetic flux $N_b$. We also derive the fluctuations and correlations of B,Q and S in QCD with three massless flavor quarks in the high-temperature limit at $eB\neq0$.

We have found that at zero temperature there does not exist any singular behavior in the quadratic fluctuations and correlations of B,Q and S in our current window of the magnetic field strength. This suggests the non-existence of the superconducting phase with $eB\lesssim3.5$ GeV$^2$. At nonzero temperatures, these fluctuations and correlations are found to possess peak structures in strong magnetic fields. This could be relevant to the singular behavior due to a plausible critical end point in the $T$-$eB$ plane of QCD phase diagram and requires further studies. By comparing our results with the HRG computations and the ideal gas limit it is found that the magnetic field fosters the QCD transition, and this is consistent with the reduction of transition temperature in magnetic fields.

Since isospin symmetry is broken in the nonzero magnetic field, we propose to investigate $(2\chi_{11}^{\rm QS}-\chi_{11}^{\rm BS})/\chi_2^{\rm S}$ and $(2\chi_{11}^{\rm BQ}-\chi_{11}^{\rm BS})/\chi_2^{\rm B}$ (cf. Fig.~\ref{fig:iso}) to detect the possible existence of a magnetic field in the late stage of heavy-ion collisions. One could check the centrality dependence of these two quantities based on the existing high energy heavy-ion experimental data as $eB$ varies in the different centrality classes. One can also do similar analyses on
$\chi_{2}^{\rm B}/\chi_{2}^{\rm S}$, $\chi_{2}^{\rm B}/\chi_{11}^{\rm QS}$ as well as $-3\chi_{11}^{BS}/\chi_2^{\rm S}$, as all these three quantities show strong $eB$ dependences at $T\lesssim169$ MeV (cf. Fig.~\ref{fig:ratio_ph}).

\section*{Acknowledgement}
We thank Frithjof Karsch and Swagato Mukherjee for useful discussions, and the HotQCD collaboration for sharing its software suite based on which the codes used in the current study for generating gauge configurations and computing the Taylor expansion coefficients in nonzero magnetic fields are developed. This work was supported by the NSFC under grant numbers 11535012, 11775096 and 11947237.
The numerical simulations have been performed on the GPU cluster in the Nuclear Science Computing Center at Central China Normal University (NSC$^3$).


\providecommand{\href}[2]{#2}\begingroup\raggedright\endgroup

\end{document}